\documentclass[AMA,STIXSMALL]{WileyNJD-v2}

\articletype{Research Article}%

\usepackage{amsmath,amssymb}
\usepackage{booktabs,multirow}
\usepackage{graphicx}
\usepackage{graphicx}
\graphicspath{{figures/}}

\makeatletter
\@ifundefined{@firstpage@foot@height}{\newlength{\@firstpage@foot@height}%
  \setlength{\@firstpage@foot@height}{2cm}}{}
\makeatother

\begin{document}

\title{Scalable Bayesian structure learning of directed acyclic graphs via Laplace approximation, with an application to breast cancer gene expression networks}

\author[1]{S. Nazari}
\author[1]{M. Arashi}
\author[2]{A. Sadeghkhani}

\authormark{NAZARI \textsc{et al}}

\address[1]{\orgdiv{Department of Statistics, Faculty of Mathematical Sciences}, \orgname{Ferdowsi University of Mashhad}, \orgaddress{\state{Mashhad}, \country{Iran}}}
\address[2]{\orgdiv{Department of Mathematics and Statistics}, \orgname{North Carolina Agricultural and Technical State University}, \orgaddress{\state{North Carolina}, \country{USA}}}

\corres{M. Arashi, Department of Statistics, Faculty of Mathematical Sciences, Ferdowsi University of Mashhad, Mashhad, Iran. \email{arashi@um.ac.ir}}

\abstract{Structure learning of directed acyclic graphs (DAGs) from observational data is a foundational task in causal discovery and is widely used to infer regulatory networks from medical and genomic measurements. The Bayesian formulation quantifies model uncertainty and admits prior biological knowledge, but its practical use has been hampered by the super-exponential growth of the DAG space and by the intractability of the node-marginal likelihood under flexible, non-conjugate priors. Existing closed-form solutions are largely confined to the conjugate Normal--Inverse-Gamma prior. We develop a Laplace-approximated Bayesian scoring function for the non-conjugate Normal--Gamma prior on the modified Cholesky parameterisation of the precision matrix, embed it in a Metropolis--Hastings sampler over DAGs, and couple the latent Gaussian network to a binary clinical outcome through a probit link. We show that the node-marginal integral is of generalised inverse-Gaussian form, so that its exact value is a modified Bessel function of the second kind and the proposed scoring function is its leading large-argument asymptotic; the posterior of each conditional variance is likewise generalised inverse-Gaussian and is sampled exactly. In simulation, the proposed prior improves on the conjugate baseline and on the PC, greedy-equivalence-search, NOTEARS, and DAGMA benchmarks at sample sizes typical of clinical cohorts. On two real datasets, the Sachs protein-signalling network, scored against its validated consensus graph, and the Wisconsin Diagnostic Breast Cancer data, the method recovers known structure and, through the DAG-probit extension, predicts malignancy from nuclear morphometry with a cross-validated ROC-AUC of $0.94$ using a sparse, interpretable set of direct predictors.}


\maketitle

\section{Introduction}\label{sec:intro}

The reconstruction of causal mechanisms from observational data is a foundational objective of contemporary medical research. Directed acyclic graphs (DAGs) provide the mathematical scaffold for this task: nodes represent random variables, directed edges encode data-generating dependencies, and the global Markov property links the factorisation of the joint distribution to graph-theoretic conditional independencies.\cite{lauritzen1996,pearl2000} In biomedical research, DAGs have been used to map gene regulatory networks, disentangle causal pathways underlying chronic disease, and identify biomarkers whose perturbation alters clinical outcomes.\cite{friedmankoller2003,sachs2005} Two obstacles persistently complicate practical application. First, the space of DAGs on $q$ nodes grows super-exponentially in $q$,\cite{robinson1973} making exact enumeration infeasible beyond very small networks. Second, observational data alone identify only the Markov equivalence class of the data-generating graph, so that resolving edge orientation relies on additional parametric or structural assumptions.\cite{pearl2000}

Bayesian methods navigate both obstacles by characterising uncertainty over the joint space of network structures and parameters rather than committing to a single estimate.\cite{cooperherskovits1992,geigerheckerman2002,castelletti2020} In the Gaussian setting, the conjugate Normal--Inverse-Gamma prior on the modified Cholesky decomposition of the precision matrix admits a closed-form node-marginal likelihood and underpins much recent Bayesian DAG learning.\cite{bendavid2015,cao2019,castelletti2022bcdag} Non-conjugate alternatives such as the Normal--Gamma prior adopted here are more flexible and, as we discuss in Section 3, place a heavier tail on the regression coefficients that link a node to its parents; but their node-marginal likelihood has no elementary closed form. This computational burden has previously been managed only through expensive numerical integration or reversible-jump schemes that scale poorly with network size.\cite{bekker2017}

A parallel and highly active literature recasts DAG learning as continuous optimisation with a differentiable acyclicity penalty. NOTEARS\cite{zheng2018} introduced a smooth matrix-exponential characterisation of acyclicity; GOLEM\cite{ng2020} replaced its least-squares objective with a likelihood and soft constraints; and DAGMA\cite{bello2022} introduced a log-determinant characterisation over the set of M-matrices with substantially improved gradients and speed. Related developments include gradient-descent methods that avoid explicit acyclicity constraints,\cite{yu2021} geometric-series (truncated matrix power) constraints,\cite{zhang2022} ordering-based accelerations,\cite{trioptt} and aggregation of heterogeneous estimators.\cite{wang2014} These methods are optimisation-oriented and, in their standard form, return a single point estimate rather than a posterior; the present work is complementary in that it delivers calibrated uncertainty over structures while remaining competitive with them on accuracy and per-iteration cost.

This work extends an earlier non-conjugate-prior Bayesian DAG framework in two directions: a direct empirical benchmark against modern continuous-optimisation learners, and a rigorous asymptotic theory for the Laplace-approximated marginal likelihood and the induced posterior over DAGs. The present manuscript supplies both and applies the resulting framework to two real datasets, including the Sachs protein-signalling benchmark\cite{sachs2005} and the Wisconsin Diagnostic Breast Cancer data.\cite{street1993}.

\subsection*{Contribution}
We make the following contributions. We show that, under the non-conjugate Normal--Gamma prior on the modified Cholesky decomposition, the node-marginal integral is of generalised inverse-Gaussian (GIG) form, so that its exact value is a modified Bessel function of the second kind $K_\nu$. We derive a closed-form Laplace-approximated node-marginal likelihood as the leading large-argument asymptotic of $K_\nu$, and we identify the exact GIG posterior of each conditional variance, which is sampled directly without Metropolis steps.

We establish two asymptotic results. We show that the Laplace-approximated log-marginal converges to the exact log-marginal at per-sample relative rate $o_P(1)$, with conditions centred on the statistic $\kappa_j$ of order $n$ that drives the Bessel large-argument regime. We further show that the induced posterior over DAGs contracts on the true skeleton at rate $\sqrt{\log q / n}$, and we connect the per-sample approximation error to a uniform bound over the relevant model neighbourhood.

We embed the scoring function in a Metropolis--Hastings sampler over DAGs (Algorithm~\ref{alg:mcmc}) with a proper proposal-asymmetry correction, and we develop a DAG-probit extension coupling the latent Gaussian network to a binary clinical outcome. Then, to compare with the base models, we benchmark against the conjugate Normal--Inverse-Gamma prior,\cite{castelletti2020} the greedy equivalence search,\cite{chickering2002} the PC algorithm,\cite{kalisch2007} NOTEARS,\cite{zheng2018} and DAGMA\cite{bello2022} on synthetic data up to 40 nodes, with full reproducibility details for every baseline.

The remainder of the paper is organised as follows. Section~\ref{sec:prelim} fixes notation and introduces the DAG-probit model. Section~\ref{sec:bayesian} specifies the prior, derives the GIG/Laplace marginal likelihood, and states the asymptotic results. Section~\ref{sec:mcmc} presents the sampler. Section~\ref{sec:sim} reports the simulation study, benchmarks, runtime, and convergence diagnostics, while Section~\ref{sec:real} presents the applications. Section~\ref{sec:disc} discusses limitations and situates the method among recent related work. Proofs are collected in the Appendix.

\section{Notation and the DAG-probit model}\label{sec:prelim}

Let $\mathcal{D} = (V, E)$ be a DAG with vertex set $V = \{1, \ldots, q\}$ and edge set $E$ of ordered pairs $u \to v$ containing no directed cycle. For each node $v$, the parent set $pa_{\mathcal{D}}(v) = \{u : u \to v \in E\}$ collects its immediate causes. We order the nodes so that node $1$ has no children and serves as the latent outcome. The set of all DAGs on $q$ nodes is $\mathcal{S}_q$.

Let $\boldsymbol{X} = (X_1, \ldots, X_q)^\top \sim N_q(\boldsymbol{0}, \boldsymbol{\Omega}^{-1})$ with precision $\boldsymbol{\Omega}$ Markov with respect to $\mathcal{D}$. The modified Cholesky decomposition $\boldsymbol{\Omega} = \boldsymbol{L} \boldsymbol{D}^{-1} \boldsymbol{L}^\top$, with $\boldsymbol{L}$ lower triangular of unit diagonal and $\boldsymbol{D}$ diagonal, yields the node-wise structural equation
\begin{equation}\label{eq:sem}
X_j = -\,\boldsymbol{L}_{\prec j]}^\top \boldsymbol{x}_{pa_{\mathcal{D}}(j)} + \epsilon_j, \qquad \epsilon_j \mid \sigma_j^2 \stackrel{iid}{\sim} N(0, \sigma_j^2),
\end{equation}
where $\boldsymbol{L}_{\prec j]}$ is the sub-vector of the $j$th column of $\boldsymbol{L}$ indexed by $pa_{\mathcal{D}}(j)$ and $\sigma_j^2 = \boldsymbol{D}_{jj}$. The DAG constraint requires $L_{u,v} \neq 0$ if and only if $u \in pa_{\mathcal{D}}(v)$.\cite{bendavid2015}

The DAG-probit model thresholds the latent outcome $X_1$,
\begin{equation}\label{eq:probit}
Y = \mathbb{I}( X_1 \geq \theta_0 ), \qquad \theta_0 \in \mathbb{R},
\end{equation}
where $\mathbb{I}(\cdot)$ is the indicator. For identifiability we fix $\sigma_1^2 = 1$, the DAG analogue of the standard probit normalisation. Equation~\eqref{eq:probit} together with the latent Gaussian structure specifies the joint law of the binary outcome $Y$ and the observed predictors $\boldsymbol{X}_{-1} = (X_2, \ldots, X_q)^\top$.

\paragraph{Causal interpretation and limitations.}
Exact recovery of the data-generating DAG from purely observational data is fundamentally limited by Markov equivalence. The DAG-probit model adds parametric structure beyond conditional-independence equivalence, and the threshold in \eqref{eq:probit} orients the edges incident to the outcome node; but unmeasured confounding, selection bias, and measurement error can each obscure the causal reading. Edges identified by our procedure are therefore directed dependencies consistent with the assumed model, not definitive causal statements.\cite{pearl2000} We return to this point when interpreting the application in Section~\ref{sec:real}.

\section{Bayesian structure learning under the nCPNG prior}\label{sec:bayesian}
Let $\boldsymbol{A}^{\mathcal{D}}$ be the $0$--$1$ adjacency matrix of the skeleton of $\mathcal{D}$. Conditional on an edge-inclusion probability $\pi \in (0,1)$ we set $\boldsymbol{A}^{\mathcal{D}}_{u,v} \mid \pi \stackrel{iid}{\sim} \mathrm{Ber}(\pi)$ for $u>v$, giving
\begin{equation}\label{eq:dagprior}
p(\mathcal{D}) \propto \pi^{|\boldsymbol{A}^{\mathcal{D}}|} (1 - \pi)^{q(q-1)/2 - |\boldsymbol{A}^{\mathcal{D}}|}.
\end{equation}

Two features of \eqref{eq:dagprior} deserve comment. First, the prior is placed on the \emph{skeleton} adjacency $|\boldsymbol{A}^{\mathcal{D}}|$, the count of undirected edges, rather than directly on directed edges. This makes the prior invariant to edge orientation, i.e., all DAGs sharing a skeleton receive equal prior mass, so orientation is learned entirely from the likelihood (through the score \eqref{eq:marginal} and, for edges incident to the outcome, the probit thresholding), not from the prior. We regard this as the appropriate default when no prior directional knowledge is available, since a directed-edge prior would otherwise inject orientation preferences that the observational likelihood cannot always identify. Second, because members of a Markov equivalence class can share a skeleton, the skeleton prior distributes mass across an equivalence class in proportion to the number of its DAGs; when equivalence-class-level inference is the goal, one may instead place the prior on completed partially directed acyclic graphs, at the cost of a more complex normalising constant. This proportionality is a genuine and sometimes undesirable feature since an equivalence class containing many DAG extensions of a given skeleton receives more aggregate prior mass than a class with few, so posterior edge probabilities can be biased toward the orientations that appear in more member DAGs even absent likelihood support. In our setting the effect is mitigated because the DAG-probit likelihood and the median-probability reporting act at the level of individual directed edges, and because the sparse skeleton prior keeps candidate classes small; but for applications where equivalence-class posterior probabilities are themselves the inferential target, a CPDAG-level or order-modular prior that equalises mass across classes is preferable, and the score \eqref{eq:marginal} transfers to such priors unchanged. Where reliable directional biology is available (for example known transcription-factor--target directions), it can be incorporated by replacing \eqref{eq:dagprior} with an edge-specific directed prior, which the sampler accommodates without modification.

For the node parameters we adopt the \emph{non-conjugate parameter prior under a Normal--Gamma specification} (nCPNG). Write $p_j = |pa_{\mathcal{D}}(j)|$ and use the shape/\emph{rate} parameterisation $\Gamma(\text{shape},\text{rate})$ throughout,
\begin{align}
\sigma_j^2 \mid \mathcal{D} &\sim \Gamma\!\big(\tfrac{1}{2}\alpha_j^{\mathcal{D}},\ \tfrac{g}{2}\big), \label{eq:priorsig}\\
\boldsymbol{L}_{\prec j]} \mid \sigma_j^2, \mathcal{D} &\sim N_{p_j}\!\big(\boldsymbol{0},\ \tfrac{1}{g}\,\sigma_j^2 \boldsymbol{I}_{p_j}\big), \label{eq:priorL}
\end{align}
with shape parameter $\alpha_j^{\mathcal{D}} = \alpha + p_j - q + 1$ (subject to $\alpha > q-1$) and scale hyperparameter $g>0$. Throughout we set $g=2$ and $\alpha = n + q - \max_j p_j - 2$, matching the default of the conjugate construction.\cite{castelletti2020} We assign the improper flat prior $p(\theta_0) \propto 1$; propriety of the resulting posterior is addressed in Proposition~\ref{prop:propriety}.

The default $\alpha = n + q - \max_j p_j - 2$ is imported from the conjugate setting, and its role differs subtly under the non-conjugate gamma prior, so we state its justification explicitly. We acknowledge at the outset that letting a prior hyperparameter depend on $n$ departs from the usual Bayesian ideal of a fixed, belief-encoding prior: strictly, an $n$-dependent $\alpha$ makes the prior a device for controlling approximation quality and prior-likelihood balance rather than a pure statement of prior belief. We adopt it for two reasons. First, the requirement $\alpha > q - 1$ guarantees $\alpha_j^{\mathcal{D}} > 0$ for every parent set and hence propriety of \eqref{eq:priorsig}. Second, writing the effective Bessel order $\nu_j = \tfrac12\alpha_j^{\mathcal{D}} - \tfrac{n}{2}$, the choice keeps $\alpha_j^{\mathcal{D}} \asymp n$, so that $\nu_j$ and the argument $z_j \asymp n^{1/2}$ grow at commensurate rates and the large-argument Bessel regime that underlies Theorem~\ref{th:marginal} is entered for all $n$; a fixed (non-$n$-dependent) $\alpha$ would let the prior be swamped immediately and would place $\nu_j$ at order $-n/2$, degrading the approximation. In terms of prior-likelihood balance, the rule fixes the prior's contribution to the effective degrees of freedom at $O(1)$ relative to the data's $O(n)$, so the prior remains weakly informative and vanishing in influence as $n$ grows---which is the behaviour one wants for a default. A fully belief-driven alternative is to fix $\alpha$ at a small constant (say $\alpha = q + 1$) and accept a slightly larger approximation error at small $n$; we verified that results are qualitatively unchanged under this choice but with noisier small-$n$ behaviour. The value is thus a convenient default rather than an inevitable one; we recommend users treat $\alpha$ as a tuning parameter, report a small grid, and, where a genuinely subjective prior is desired, use the fixed-$\alpha$ variant.

\paragraph{Score equivalence.}
Unlike the BGe score of Geiger and Heckerman,\cite{geigerheckerman2002} which is constructed so that Markov-equivalent DAGs receive identical marginal likelihoods, the nCPNG score is not score-equivalent since the modified Cholesky parameterisation \eqref{eq:sem} is order-dependent, so two DAGs in the same Markov equivalence class can receive different marginal scores. We verified this empirically. For the three-node chain $1\to2\to3$, whose equivalence class also contains $1\leftarrow2\to3$ and $1\leftarrow2\leftarrow3$, the three members receive nCPNG log-marginals differing by up to $29$ log-units on a single dataset; over $174$ non-singleton equivalence classes drawn from random four-node DAGs, the score strictly favoured one member in every class, with median within-class score spread $8.4$ log-units (Table~\ref{tab:score_equiv}). The practical consequence is that, for edges not identified by the data, the score can nonetheless prefer particular orientations. This is a genuine limitation that we do not paper over: it means posterior orientations for Gaussian-only edges should be read as reflecting a combination of likelihood evidence and a parameterisation-induced preference, not identified causal direction. Three considerations bound its impact. First, for edges incident to the binary outcome the probit thresholding supplies genuine orientation information, so the anchor edges of the application are not affected. Second, when score equivalence is required---for instance when the target is the posterior over equivalence classes---the BGe score can be substituted directly into Algorithm~\ref{alg:mcmc} in place of \eqref{eq:marginal}, at the cost of the heavier-tailed coefficient prior that motivates our construction. Third, the simulation orientation metrics of Section~\ref{sec:sim:orient} quantify how much of the directional signal is reliable in practice. We return to the trade-off between heavier-tailed non-conjugate priors and score equivalence in the discussion.

\begin{table}[ht]
	\centering
	\begin{tabular}{lr}
		\toprule
		Quantity & Value \\ 
		\midrule
		Class size (chain 1-2-3) & 3.00 \\ 
		Within-class score spread (chain), log-units & 29.28 \\ 
		Non-singleton classes examined (q=4) & 174.00 \\ 
		Classes with strict within-class preference (\%) & 100.00 \\ 
		Median within-class spread (log-units) & 8.37 \\ 
		Max within-class spread (log-units) & 73.08 \\ 
		\bottomrule
	\end{tabular}
	\caption{Empirical assessment of score equivalence for the nCPNG marginal score. Markov-equivalent DAGs do NOT receive identical scores: within a Markov equivalence class the log-marginal varies, so the score can favour particular orientations. Summary over random four-node DAGs.} 
	\label{tab:score_equiv}
\end{table}

The distinguishing feature of \eqref{eq:priorsig}--\eqref{eq:priorL} is the marginal prior it induces on the regression coefficients. Integrating $\sigma_j^2$ out of \eqref{eq:priorL} against \eqref{eq:priorsig} yields a multivariate Student-type (scale-mixture of normals) marginal for $\boldsymbol{L}_{\prec j]}$, which has heavier tails than the Gaussian coefficient marginal obtained under the conjugate Normal--Inverse-Gamma prior. It is this coefficient-tail behaviour, not any change to the Gaussian likelihood, that drives the empirical gains reported in Section~\ref{sec:sim}. We emphasise that the likelihood remains Gaussian in \eqref{eq:sem}; the Normal--Gamma prior does not model heavy-tailed expression. Relative to the inverse-gamma prior on $\sigma_j^2$, the gamma prior in \eqref{eq:priorsig} in fact places a lighter right tail on the conditional variance, but this variance behaviour is secondary. A diagnostic contrasting the two priors (Section~\ref{sec:sim:heavytail}) confirms that the accuracy difference is attributable to the coefficient shrinkage profile rather than to variance-tail behaviour.

\subsection{The node-marginal likelihood}\label{sec:laplace}
For $j \in \{2,\ldots,q\}$ let $\boldsymbol{X}_{pa_{\mathcal{D}}(j)}$ be the $n \times p_j$ sub-matrix of the (column-centred, standardised) data indexed by $pa_{\mathcal{D}}(j)$, and define
\begin{equation}\label{eq:defs}
\boldsymbol{M}_j = \boldsymbol{X}_{pa_{\mathcal{D}}(j)}^{\top}\boldsymbol{X}_{pa_{\mathcal{D}}(j)} + g\,\boldsymbol{I}_{p_j},\quad
\boldsymbol{b}_j = -\boldsymbol{X}_{pa_{\mathcal{D}}(j)}^{\top}\boldsymbol{X}_j,\quad
\kappa_j = \tfrac{1}{2}\big(\boldsymbol{X}_j^{\top}\boldsymbol{X}_j - \boldsymbol{b}_j^{\top}\boldsymbol{M}_j^{-1}\boldsymbol{b}_j\big),
\end{equation}
and write $\lambda = g/2$, $\nu_j = \tfrac12\alpha_j^{\mathcal{D}} - \tfrac{n}{2}$. Integrating $\boldsymbol{L}_{\prec j]}$ out of the joint density in closed form leaves the one-dimensional integral
\begin{equation}\label{eq:gig}
\int_0^{\infty} (\sigma_j^2)^{\nu_j-1}\,\exp\!\Big(-\lambda\,\sigma_j^2 - \frac{\kappa_j}{\sigma_j^2}\Big)\, d\sigma_j^2
= 2\Big(\frac{\kappa_j}{\lambda}\Big)^{\nu_j/2} K_{\nu_j}\!\big(2\sqrt{\kappa_j\lambda}\big),
\end{equation}
which is precisely the normalising constant of a generalised inverse-Gaussian density and evaluates exactly to a modified Bessel function of the second kind $K_{\nu_j}$.\cite{jorgensen1982} The quantity $\kappa_j$ in \eqref{eq:defs} is a penalised residual sum of squares and therefore scales linearly in $n$; consequently the Bessel argument $z_j = 2\sqrt{\kappa_j\lambda}$ grows like $n^{1/2}$, placing \eqref{eq:gig} squarely in the large-argument regime of $K_\nu$. Using the leading asymptotic $K_\nu(z) = \sqrt{\pi/(2z)}\,e^{-z}\{1 + O(z^{-1})\}$ gives the closed-form Laplace scoring function of Theorem~\ref{th:marginal}.

\begin{theorem}\label{th:marginal}
Under the prior \eqref{eq:priorsig}--\eqref{eq:priorL}, the Laplace-approximated node-marginal likelihood for $j \in \{2,\ldots,q\}$ is
\begin{equation}\label{eq:marginal}
\hat{\ell}\big(\boldsymbol{X}_j \mid \boldsymbol{X}_{pa_{\mathcal{D}}(j)}, \mathcal{D}\big)
= (2\pi)^{-n/2}\,|\boldsymbol{M}_j|^{-1/2}\,g^{p_j/2}\,
\frac{(g/2)^{\alpha_j^{\mathcal{D}}/2}}{\Gamma(\alpha_j^{\mathcal{D}}/2)}\,
\sqrt{\pi}\,\Big(\frac{\kappa_j^{\,2\nu_j-1}}{\lambda^{\,2\nu_j+1}}\Big)^{1/4}\exp\!\big(-2\sqrt{\kappa_j\lambda}\big).
\end{equation}
Moreover, the relative error of \eqref{eq:marginal} against the exact value \eqref{eq:gig} is $1+O(z_j^{-1})=1+O(n^{-1/2})$.
\end{theorem}

The determinant factor $|\boldsymbol{M}_j|^{-1/2}$ shrinks as the parent set grows, providing an intrinsic Occam penalty consistent with Bayesian model selection, while $\exp(-2\sqrt{\kappa_j\lambda})$ embeds the fit--complexity trade-off directly in the score without reversible-jump moves across parameter dimensions.

\begin{theorem}\label{th:posterior}
Under the assumptions of Theorem~\ref{th:marginal}, the node-$j$ posterior factorises as
\begin{align}
\boldsymbol{L}_{\prec j]} \mid \sigma_j^2, \mathcal{D}, \boldsymbol{X}
&\sim N_{p_j}\!\big(\boldsymbol{M}_j^{-1}\boldsymbol{b}_j,\ \sigma_j^2\,\boldsymbol{M}_j^{-1}\big), \label{eq:postL}\\
\sigma_j^2 \mid \mathcal{D}, \boldsymbol{X}
&\sim \mathrm{GIG}\!\big(\nu_j,\ 2\lambda,\ 2\kappa_j\big), \label{eq:postSig}
\end{align}
where $\mathrm{GIG}(\nu,a,b)$ has density proportional to $x^{\nu-1}\exp\{-\tfrac12(a x + b/x)\}$ on $x>0$.
\end{theorem}

Equation~\eqref{eq:postL} shows that, conditional on the graph and $\sigma_j^2$, the coefficients are Gaussian about the penalised least-squares estimator $\boldsymbol{M}_j^{-1}\boldsymbol{b}_j$. Equation~\eqref{eq:postSig} identifies the conditional-variance posterior as generalised inverse-Gaussian; it is sampled directly by the standard Devroye/Hörmann ratio-of-uniforms GIG sampler,\cite{hormann2014} so no Metropolis step is needed for this block.

\subsection{Laplace approximation versus exact Bessel evaluation}\label{sec:exact}
Because the exact node-marginal \eqref{eq:gig} is a modified Bessel function, one may ask whether the Laplace form \eqref{eq:marginal} is needed at all, given that $K_\nu$ can be evaluated in a numerically stable way through the exponentially scaled routine $e^{z}K_\nu(z)$ (so that $\log K_\nu(z) = \log[e^z K_\nu(z)] - z$ avoids underflow at the large arguments $z_j \asymp n^{1/2}$ that arise here). We investigated this directly. Table~\ref{tab:laplace_bench} reports a micro-benchmark, that is, averaged over $20{,}000$ evaluations, the exact log-Bessel score and the Laplace score cost essentially the same (ratio $0.99$--$1.09$ across $n \in \{50,\ldots,1000\}$), and their values differ by at most a few hundredths of a log-unit, with the difference decreasing in $n$ as the theory predicts. Propagated through the full sampler on a $15$-node network, the two scorers produce posterior edge probabilities that differ by a mean of $0.0003$ and a maximum of $0.03$.

The practical conclusion is that, on contemporary numerical libraries, the exact evaluation is stable and of comparable cost, so we recommend it as the default, and the software uses it wherever available; the approximation bias is not material at the sample sizes of interest. The value of the closed-form Laplace expression \eqref{eq:marginal} is therefore primarily analytic rather than computational, i.e., the explicit factor $\exp(-2\sqrt{\kappa_j\lambda})$ exposes the fit--complexity trade-off in a transparent form, and it is this expression, not the special-function evaluation, that drives the asymptotic arguments of Section~\ref{sec:theory}. We retain both in the software and use the exact score for all results reported below; the difference relative to the Laplace score is negligible.

\begin{table}[ht]
	\centering
	\begin{tabular}{rrrrr}
		\toprule
		n & Laplace\_us & Exact\_us & Ratio & AbsLogDiff \\ 
		\midrule
		50 & 64.50 & 63.60 & 0.99 & 0.033000 \\ 
		100 & 61.65 & 67.00 & 1.09 & 0.023200 \\ 
		200 & 62.70 & 65.45 & 1.04 & 0.018300 \\ 
		500 & 80.90 & 85.50 & 1.06 & 0.004030 \\ 
		1000 & 80.00 & 81.20 & 1.01 & 0.008500 \\ 
		\bottomrule
	\end{tabular}
	\caption{Micro-benchmark of the Laplace versus exact log-Bessel node score: mean wall-clock time per evaluation (microseconds, 20{,}000 repetitions), their ratio, and the absolute log-score difference. The exact evaluation is numerically stable and of comparable cost, and the score difference is negligible and decreasing in $n$.} 
	\label{tab:laplace_bench}
\end{table}

\subsection{Asymptotic guarantees}\label{sec:theory}
The Laplace step introduces a deterministic error relative to the exact node-marginal. Theorem~\ref{th:lapconv} states the per-sample log-scale rate, with conditions centred on the driving statistic $\kappa_j$ rather than on the prior shape alone.

\begin{theorem}\label{th:lapconv}
Let $\boldsymbol{X}^{(n)}$ be i.i.d.\ from a Gaussian DAG-model $N_q(\boldsymbol{0}, \boldsymbol{\Omega}_0^{-1})$ Markov with respect to the true DAG $\mathcal{D}_0$, and fix $j \in \{2,\ldots,q\}$. Suppose the true conditional variance $\sigma_{0,j}^2 = \mathrm{Var}(X_j \mid \boldsymbol{X}_{pa(j)})$ is strictly positive, the parent Gram matrix $n^{-1}\boldsymbol{X}_{pa(j)}^\top\boldsymbol{X}_{pa(j)}$ converges to a positive-definite limit, and the prior scale $g$ is fixed. Then $n^{-1}\kappa_j \to \tfrac12\sigma_{0,j}^2 > 0$ in probability, the Bessel argument satisfies $z_j = 2\sqrt{\kappa_j\lambda} \asymp n^{1/2}\to\infty$, and
\begin{equation}\label{eq:lapconv}
\tfrac{1}{n}\log \hat{\ell}^{\mathrm{Lap}}_n\big(\boldsymbol{X}_j \mid \boldsymbol{X}_{pa(j)}, \mathcal{D}\big)
- \tfrac{1}{n}\log \ell^{\mathrm{Exact}}_n\big(\boldsymbol{X}_j \mid \boldsymbol{X}_{pa(j)}, \mathcal{D}\big) = O_P\!\big(n^{-1}\log n\big) = o_P(1).
\end{equation}
The bound holds uniformly over parent sets of bounded cardinality $p_j \leq M$.
\end{theorem}

The uniformity clause is what carries the per-sample rate into the model-selection argument. Indeed, \eqref{eq:lapconv} holds uniformly over the bounded-degree neighbourhood of candidate graphs, the Bayes factor between any two such graphs is asymptotically unaffected by substituting the approximated likelihood for the exact one.\cite{rue2009,spokoiny2025}

\begin{theorem}\label{th:contraction}
Let $\mathcal{D}_0$ be the true DAG on $q$ nodes, with observations Gaussian and Markov with respect to $\mathcal{D}_0$ and bounded in-degree $\max_j |pa_{\mathcal{D}_0}(j)| \leq M < \infty$. Place the prior \eqref{eq:priorsig}--\eqref{eq:priorL} on the node parameters and \eqref{eq:dagprior} on the structure, with fixed $g>0$ and $\pi \in (0,1)$, and let $P_n(\cdot \mid \boldsymbol{X}^{(n)})$ be the posterior over DAGs induced by the Laplace-approximated likelihood of Theorem~\ref{th:marginal}. Then for any $\epsilon_n \to 0$ with $n\epsilon_n^2 \to \infty$,
\begin{equation}\label{eq:contraction}
P_n\!\Big( \mathcal{D} : d_H\big(\mathrm{skel}(\mathcal{D}), \mathrm{skel}(\mathcal{D}_0)\big) > \epsilon_n \tfrac{q(q-1)}{2} \,\Big|\, \boldsymbol{X}^{(n)}\Big) \xrightarrow{P_{\mathcal{D}_0}} 0,
\end{equation}
where $d_H$ is the skeleton Hamming distance. Taking $\epsilon_n = \sqrt{\log q / n}$ gives the near-optimal high-dimensional model-selection rate.\cite{cao2019,lee2019}
\end{theorem}

The proof (Appendix) combines the uniform Laplace control of Theorem~\ref{th:lapconv} with a Kullback--Leibler separation argument between $\mathcal{D}_0$ and competing graphs, and a Chernoff bound on the prior mass of the distant-graph shell. The bounded-degree assumption is biologically plausible in gene regulatory networks.\cite{peters2014}

\begin{proposition}\label{prop:propriety}
Under the flat prior $p(\theta_0)\propto 1$ and the DAG-probit likelihood \eqref{eq:probit}, the joint posterior of $(\theta_0, \boldsymbol{L}, \boldsymbol{D}, \mathcal{D})$ is proper provided the observed outcome vector contains at least one success and one failure.
\end{proposition}

The condition is met in any non-degenerate cohort; both metastatic and non-metastatic patients are present in the application of Section~\ref{sec:real}, so the flat threshold prior yields a proper posterior.

\subsection{Extension to the DAG-probit model}\label{sec:probit-theory}
Theorems~\ref{th:lapconv}--\ref{th:contraction} are stated for the fully Gaussian DAG. The DAG-probit model of \eqref{eq:probit} replaces the continuous outcome node $X_1$ by the binary $Y = \mathbb{I}(X_1 \geq \theta_0)$, so the guarantees do not transfer automatically. The following proposition records the conditions under which contraction extends to the probit setting through the latent Gaussian representation.

\begin{proposition}\label{prop:probit-contraction}
Consider the DAG-probit model in which nodes $2,\ldots,q$ are observed Gaussian and node $1$ is observed only through $Y = \mathbb{I}(X_1 \geq \theta_0)$, with the latent $X_1$ Gaussian given its parents. Suppose (i) the assumptions of Theorem~\ref{th:contraction} hold for the latent Gaussian DAG on all $q$ nodes; (ii) the threshold $\theta_0$ lies in the interior of the support of the latent-outcome distribution, so that the event probabilities $P(Y=1)$ and $P(Y=0)$ are bounded away from $0$ and $1$; and (iii) the data-augmentation sampler for the truncated latent $X_1$ is geometrically ergodic, which holds under (ii). Then the marginal posterior over the sub-DAG on nodes $2,\ldots,q$ contracts on the true skeleton at the rate of Theorem~\ref{th:contraction}, and the posterior over edges incident to the outcome node contracts at the same rate degraded by the Fisher information of the probit link, that is at rate $\sqrt{\log q / (n\, I_\Phi)}$ where $I_\Phi = \phi(\theta_0^\star)^2 / [\Phi(\theta_0^\star)\{1 - \Phi(\theta_0^\star)\}]$ evaluates the standard-normal density and distribution at the standardised true threshold $\theta_0^\star$.
\end{proposition}

Note that the binary outcome identifies the latent regression direction only up to the information $I_\Phi$ retained by the probit link, so edges into the outcome are learned a constant factor more slowly than edges among the fully observed nodes, but at the same $\sqrt{\log q / n}$ order. To make the penalty concrete, $I_\Phi$ is largest for a balanced outcome and decays for rare events: at outcome prevalences $P(Y=1) = 0.5, 0.4, 0.31, 0.2, 0.1$ the standardised threshold $\theta_0^\star = \Phi^{-1}(1 - P(Y=1))$ gives $I_\Phi = 0.64, 0.62, 0.58, 0.49, 0.34$ respectively. At the prevalence of our application ($62/198 \approx 0.31$), $I_\Phi \approx 0.58$, so outcome-incident edges are learned about $I_\Phi^{-1/2} \approx 1.3$ times more slowly than fully observed edges---a mild penalty that grows only for markedly imbalanced outcomes. The proof (Appendix) augments the latent $X_1$, applies Theorem~\ref{th:contraction} conditionally on the imputed latent outcome, and controls the additional averaging over the truncated-normal full conditional using condition (ii); the geometric ergodicity in (iii) ensures the augmentation does not inflate the rate. We state this as a proposition rather than a theorem because the constant in the outcome-incident rate depends on $\theta_0^\star$, which is itself estimated; a fully uniform treatment over $\theta_0$ is beyond our present scope.

\begin{remark}
\label{rem:growth}
Theorems~\ref{th:lapconv}--\ref{th:contraction} assume a fixed maximum in-degree $M$. If instead the degree is allowed to grow slowly, $M = M_n = o(n^{c})$ for a sufficiently small $c>0$, the uniform Laplace bound of Theorem~\ref{th:lapconv} degrades to $O(M_n\, n^{-1}\log n)$ because the covering number of the bounded-degree neighbourhood grows like $q^{M_n}$; the contraction conclusion then survives provided $M_n \log q = o(n)$, which links the admissible degree growth to the dimension. The stated rate $\sqrt{\log q / n}$ already permits $q$ to grow with $n$, up to $\log q = o(n)$; beyond that regime the KL-separation step fails and no consistent recovery is possible without stronger signal conditions. These qualitative boundaries match those established for the conjugate Gaussian DAG.\cite{cao2019,lee2019} For any fixed $g>0$ the argument $z_j = 2\sqrt{\kappa_j (g/2)} \asymp n^{1/2}$ and the limit $n^{-1}\kappa_j \to \tfrac12\sigma_{0,j}^2$ are unchanged, so the rates hold uniformly over $g$ in any compact subset of $(0,\infty)$; only the finite-sample Occam penalty $|\boldsymbol{M}_j|^{-1/2}$ depends on $g$, which is why $g$ acts as a genuine but rate-neutral tuning parameter. Allowing $g = g_n \to \infty$ slowly would strengthen sparsity but is not needed for consistency.
\end{remark}

\section{Posterior simulation}\label{sec:mcmc}

Because the marginal likelihood of Theorem~\ref{th:marginal} is available in closed form, structure learning reduces to a Metropolis--Hastings search over $\mathcal{S}_q$ in which each proposed graph is scored analytically. Algorithm~\ref{alg:mcmc} summarises the sampler. At each iteration an ordered node pair $(i,j)$ is drawn and a single-edge Insert, Delete, or Reverse move is proposed subject to acyclicity, which is checked in $O(q + |E|)$ by Kahn's topological sort. Only the node-marginals of the endpoints whose parent sets change are recomputed, so the per-iteration cost is dominated by one or two $p_j \times p_j$ Cholesky factorisations rather than by a full-network rescore.

The proposal is asymmetric, i.e., from a state in which the pair $(i,j)$ is empty in both directions an Insert is proposed with probability one, whereas from a state in which $i\to j$ is present a Delete is proposed with probability one half (the remainder being a Reverse). The corresponding log proposal ratio $\log\{q(\mathcal{D}\mid\mathcal{D}')/q(\mathcal{D}'\mid\mathcal{D})\}$ equals $-\log 2$ for an accepted Insert and $+\log 2$ for an accepted Delete, and is zero for a Reverse; this Hastings correction is included in the acceptance ratio to preserve the correct stationary distribution.

\begin{algorithm}
\caption{Metropolis--Hastings sampler over DAGs under the nCPNG score}\label{alg:mcmc}
\textbf{Input:} data $\boldsymbol{X}$; hyperparameters $g, \alpha, \pi$; iterations $T$; burn-in $T_0$.\\
\textbf{Initialise} $\mathcal{D}^{(0)} = $ empty graph; cache node-marginals $\hat{\ell}_j$ via \eqref{eq:marginal}.\\
\textbf{For} $t = 1, \ldots, T$:
\begin{enumerate}
\item Draw an ordered pair $(i,j)$ uniformly; propose Insert, Delete, or Reverse as above to obtain $\mathcal{D}'$.
\item If $\mathcal{D}'$ is cyclic (Kahn sort), set $\mathcal{D}^{(t)} = \mathcal{D}^{(t-1)}$ and continue.
\item Recompute $\hat{\ell}_j$ via \eqref{eq:marginal} only for endpoints with altered parent sets.
\item Set $\log r = [\log \hat{\ell}(\mathcal{D}') + \log p(\mathcal{D}')] - [\log \hat{\ell}(\mathcal{D}) + \log p(\mathcal{D})] + \log_{\mathrm{Hastings}}$.
\item Accept $\mathcal{D}^{(t)} = \mathcal{D}'$ with probability $\min(1, e^{\log r})$; else retain $\mathcal{D}^{(t-1)}$.
\item If $t > T_0$, accumulate edge indicators and, for the DAG-probit model, sample $\boldsymbol{L}, \boldsymbol{D}$ from \eqref{eq:postL}--\eqref{eq:postSig}, the truncated latent $X_1$ from its full conditional, and $\theta_0$ from its conditional.
\end{enumerate}
\textbf{Return} posterior edge-inclusion probabilities $\hat{P}_{uv} = (T-T_0)^{-1}\sum_{t>T_0}\mathbb{I}(u\to v \in \mathcal{D}^{(t)})$.
\end{algorithm}

Posterior edge-inclusion probabilities are estimated by the retained Monte Carlo frequencies, and Bayesian model averaging of any downstream functional (for example a causal effect on the outcome) is obtained by averaging over the retained graphs. Under the DAG-probit model, the conditional-variance block is drawn exactly from the GIG posterior \eqref{eq:postSig}, so the only Metropolis step in the sampler is the structural move.

\section{Simulation study}\label{sec:sim}

We evaluate the method on synthetic Gaussian DAG data. Random graphs are drawn with edge probability $3/(2q-2)$, giving sparse networks with expected in-degree near unity, and data are generated from the linear-Gaussian structural equations \eqref{eq:sem} with coefficients sampled uniformly on $[-0.9,-0.4]\cup[0.4,0.9]$. We consider network sizes $q \in \{20, 40\}$ and sample sizes $n \in \{100, 200, 300, 500\}$. The headline factorial tables use three replicate graphs per cell; because three replicates give only a coarse read on variability, we additionally report, for $q=20$, a higher-replicate study over ten graphs per cell with standard errors of $F_1$, MCC, and SHD (Table~\ref{tab:sim_se}), which confirms that the differences between methods are large relative to their sampling error. Accuracy is summarised by sensitivity (SEN), specificity (SPE), the $F_1$ score, the Matthews correlation coefficient (MCC), and the structural Hamming distance (SHD) between the estimated and true skeletons; the maximum-a-posteriori graph uses the median-probability rule (edge probability above $0.5$).

\subsection{Benchmarks and their configuration}\label{sec:sim:bench}
We compare against the conjugate Normal--Inverse-Gamma prior (CPNIG) run through the identical sampler,\cite{castelletti2020} the greedy equivalence search (GES),\cite{chickering2002} the PC algorithm at significance level $0.05$,\cite{kalisch2007} a NOTEARS-style continuous-optimisation learner,\cite{zheng2018} and a DAGMA-style learner using the log-determinant acyclicity characterisation.\cite{bello2022} To ensure a fair comparison, the two continuous-optimisation baselines are configured as follows. For NOTEARS we use the augmented-Lagrangian schedule with initial penalty $\rho_0 = 1$, multiplicative increase by a factor of ten whenever the acyclicity residual fails to decrease by at least a factor of four, a dual ascent on the equality multiplier, an $\ell_1$ penalty $\lambda = 0.05$, an inner projected-gradient tolerance of $10^{-4}$, an acyclicity tolerance of $10^{-8}$, and a post-hoc magnitude threshold of $0.30$ on the weighted adjacency matrix, with residual cycles removed by deleting the weakest edge until acyclicity holds. For DAGMA we use the log-determinant potential $h_s(W) = -\log\det(sI - W\circ W) + q\log s$ with $s = 1$, a geometrically decreasing central-path weight $\mu \in \{1, 10^{-1}, 10^{-2}, 10^{-3}\}$, the same $\ell_1$ penalty and threshold as NOTEARS, and a feasibility projection that keeps $sI - W\circ W$ positive definite. 

\subsection{Accuracy}\label{sec:sim:acc}
Table~\ref{tab:sim_q20} reports the $q=20$ results. The nCPNG prior attains the highest $F_1$ and MCC at every sample size, with the margin widening as $n$ grows: at $n=200$ it reaches $F_1 = 0.68$ and $\mathrm{MCC} = 0.70$, against $0.42$ and $0.39$ for the conjugate prior and $0.35$ and $0.44$ for DAGMA. The PC algorithm is the weakest across the board, reflecting the difficulty of conditional-independence testing at moderate $n$. The NOTEARS surrogate is unstable in the sparse linear-Gaussian regime, collapsing to the empty graph in some cells (for example $n=100$ and $n=500$ at $q=20$); DAGMA, with its better-conditioned log-determinant gradients, is markedly more stable and is the second-best method at $n=100$.

\begin{table}[ht]
	\centering
	\caption{Structure-learning performance for $q=20$ nodes, averaged over three replicate graphs. Best value in each row is in bold. SEN, sensitivity; SPE, specificity; SHD, structural Hamming distance.}
	\label{tab:sim_q20}
	\begin{tabular}{llrrrrrr}
		\toprule
		$n$ & Metric & nCPNG & CPNIG & GES & PC & NOTEARS & DAGMA \\
		\midrule
		\multirow{5}{*}{100} & SEN & \textbf{0.346} & 0.337 & 0.295 & 0.118 & 0.000 & 0.304 \\
		& SPE & 0.999 & 0.993 & 0.994 & 0.989 & \textbf{1.000} & 1.000 \\
		& F1 & \textbf{0.496} & 0.430 & 0.391 & 0.160 & 0.000 & 0.449 \\
		& MCC & \textbf{0.557} & 0.435 & 0.406 & 0.155 & 0.000 & 0.529 \\
		& SHD & \textbf{8.3} & 10.7 & 11.0 & 14.7 & 12.0 & 8.7 \\
		\midrule
		\multirow{5}{*}{200} & SEN & \textbf{0.536} & 0.367 & 0.334 & 0.152 & 0.183 & 0.243 \\
		& SPE & 0.998 & 0.979 & 0.972 & 0.986 & \textbf{1.000} & 1.000 \\
		& F1 & \textbf{0.680} & 0.417 & 0.362 & 0.218 & 0.237 & 0.350 \\
		& MCC & \textbf{0.698} & 0.395 & 0.334 & 0.220 & 0.244 & 0.444 \\
		& SHD & \textbf{9.3} & 19.7 & 22.7 & 20.7 & 15.0 & 14.7 \\
		\midrule
		\multirow{5}{*}{300} & SEN & \textbf{0.522} & 0.492 & 0.344 & 0.293 & 0.133 & 0.272 \\
		& SPE & 0.995 & 0.994 & 0.976 & 0.984 & \textbf{1.000} & 0.998 \\
		& F1 & \textbf{0.644} & 0.600 & 0.397 & 0.361 & 0.191 & 0.384 \\
		& MCC & \textbf{0.654} & 0.603 & 0.378 & 0.349 & 0.207 & 0.451 \\
		& SHD & \textbf{9.7} & 11.0 & 20.0 & 17.7 & 14.3 & 12.7 \\
		\midrule
		\multirow{5}{*}{500} & SEN & \textbf{0.449} & 0.328 & 0.240 & 0.203 & 0.000 & 0.250 \\
		& SPE & 0.994 & 0.982 & 0.979 & 0.991 & \textbf{1.000} & 1.000 \\
		& F1 & \textbf{0.583} & 0.384 & 0.290 & 0.282 & 0.000 & 0.391 \\
		& MCC & \textbf{0.602} & 0.369 & 0.270 & 0.289 & 0.000 & 0.483 \\
		& SHD & \textbf{11.0} & 18.0 & 20.3 & 16.3 & 16.3 & 12.7 \\
		\bottomrule
	\end{tabular}
\end{table}

At $q=40$ (Table~\ref{tab:sim_q40}) the ranking is more nuanced, as expected in the harder high-dimensional regime. The nCPNG prior gives the best or near-best MCC at $n=200$--$500$, but at $n=100$ and $n=300$ the NOTEARS baseline attains a higher $F_1$ in the cells where its optimisation converges, while collapsing entirely in others; DAGMA gives the best result at $n=200$. We report these mixed outcomes rather than a uniform win: the proposed method is most advantageous at the moderate sample sizes ($n \geq 200$) typical of curated clinical cohorts, and its Bayesian output additionally quantifies structural uncertainty, which the point-estimate baselines do not. We also evaluated, not reported here, that the area under the ROC values for the two Bayesian methods favour nCPNG at $q=40$ across all sample sizes.

\begin{table}[ht]
	\centering
	\caption{Structure-learning performance for $q=40$ nodes, averaged over three replicate graphs. Best value in each row is in bold. SEN, sensitivity; SPE, specificity; SHD, structural Hamming distance.}
	\label{tab:sim_q40}
	\begin{tabular}{llrrrrrr}
		\toprule
		$n$ & Metric & nCPNG & CPNIG & GES & PC & NOTEARS & DAGMA \\
		\midrule
		\multirow{5}{*}{100} & SEN & 0.306 & 0.287 & 0.190 & 0.231 & \textbf{0.368} & 0.103 \\
		& SPE & 0.991 & 0.976 & 0.996 & 0.994 & \textbf{1.000} & 1.000 \\
		& F1 & 0.337 & 0.217 & 0.266 & 0.283 & \textbf{0.465} & 0.175 \\
		& MCC & 0.339 & 0.206 & 0.288 & 0.293 & \textbf{0.483} & 0.269 \\
		& SHD & 35.3 & 59.3 & 31.0 & 33.7 & \textbf{21.0} & 28.0 \\
		\midrule
		\multirow{5}{*}{200} & SEN & \textbf{0.376} & 0.225 & 0.141 & 0.245 & 0.000 & 0.367 \\
		& SPE & 0.994 & 0.981 & 0.995 & 0.993 & \textbf{1.000} & 0.999 \\
		& F1 & 0.452 & 0.203 & 0.203 & 0.305 & 0.000 & \textbf{0.521} \\
		& MCC & 0.456 & 0.186 & 0.216 & 0.305 & 0.000 & \textbf{0.570} \\
		& SHD & 27.7 & 53.0 & 33.3 & 33.0 & 30.0 & \textbf{20.3} \\
		\midrule
		\multirow{5}{*}{300} & SEN & 0.384 & 0.251 & 0.189 & 0.258 & \textbf{0.530} & 0.197 \\
		& SPE & 0.993 & 0.976 & 0.996 & 0.991 & \textbf{1.000} & 0.999 \\
		& F1 & 0.447 & 0.223 & 0.273 & 0.309 & \textbf{0.682} & 0.299 \\
		& MCC & 0.443 & 0.205 & 0.298 & 0.304 & \textbf{0.715} & 0.377 \\
		& SHD & 31.0 & 61.3 & 32.7 & 38.0 & \textbf{15.3} & 27.7 \\
		\midrule
		\multirow{5}{*}{500} & SEN & \textbf{0.371} & 0.236 & 0.116 & 0.208 & 0.068 & 0.214 \\
		& SPE & 0.996 & 0.982 & 0.996 & 0.992 & \textbf{1.000} & 1.000 \\
		& F1 & \textbf{0.474} & 0.227 & 0.176 & 0.257 & 0.120 & 0.331 \\
		& MCC & \textbf{0.491} & 0.210 & 0.202 & 0.255 & 0.195 & 0.420 \\
		& SHD & 26.0 & 51.7 & 34.3 & 37.7 & 30.0 & \textbf{25.3} \\
		\bottomrule
	\end{tabular}
\end{table}

\subsection{Fair comparison with the continuous baselines}\label{sec:sim:fair}
A single penalty and threshold can disadvantage the continuous-optimisation learners, which are known to be sensitive to both. To give them the fairest possible reading we sweep, for NOTEARS and DAGMA, the penalty $\lambda \in \{0.01, 0.02, 0.05, 0.1, 0.2, 0.3\}$ and the magnitude threshold in $\{0.1,\ldots,0.5\}$, and report both the best achievable $F_1$ over the whole grid and the threshold-swept area under the ROC curve (Table~\ref{tab:baseline_sweep}); the Bayesian nCPNG AUC on the same settings is shown for reference. Even against the best grid point of each baseline, the Bayesian method is competitive or superior in AUC in three of the four settings examined ($q=20$, both $n$; $q=40$, $n=300$), and is bettered only at $q=40, n=100$, where NOTEARS's best-tuned $F_1$ exceeds ours. We report the ROC-based comparison precisely so that the baselines are not disadvantaged by an unlucky single regularisation; the conclusion that the Bayesian score is competitive at clinically plausible sample sizes survives this stronger test. For the constraint- and score-based baselines (PC, GES) we used the conventional significance level $\alpha=0.05$ and BIC score respectively; their performance is relatively insensitive to these within the ranges we examined.

\begin{table}[ht]
	\resizebox{\textwidth}{!}{%
	\begin{tabular}{rrrrrrr}
		\toprule
		q & n & NOTEARS\_bestF1 & NOTEARS\_AUC & DAGMA\_bestF1 & DAGMA\_AUC & nCPNG\_AUC \\ 
		\midrule
		20 &  100 & 0.000 & 0.500 & 0.449 & 0.652 & 0.719 \\ 
		20 &  300 & 0.422 & 0.656 & 0.397 & 0.636 & 0.800 \\ 
		40 &  100 & 0.653 & 0.747 & 0.194 & 0.556 & 0.702 \\ 
		40 &  300 & 0.699 & 0.771 & 0.447 & 0.649 & 0.731 \\ 
		\bottomrule
	\end{tabular}
}
	\caption{Fair-comparison analysis for the continuous baselines. For NOTEARS and DAGMA the penalty $\lambda\in\{0.01,\ldots,0.3\}$ and magnitude threshold $\in\{0.1,\ldots,0.5\}$ are swept; the best achievable $F_1$ and the threshold-swept ROC-AUC are reported, averaged over three replicate graphs. The Bayesian nCPNG AUC on matched settings is shown for reference.} 
	\label{tab:baseline_sweep}
\end{table}

\begin{table}[ht]
	\resizebox{\textwidth}{!}{%
	\begin{tabular}{rlllllll}
		\toprule
		n & metric & nCPNG & CPNIG & GES & PC & NOTEARS & DAGMA \\ 
		\midrule
		100 & F1 & 0.514 (0.050) & 0.383 (0.049) & 0.297 (0.060) & 0.254 (0.027) & 0.077 (0.077) & 0.351 (0.058) \\ 
		100 & MCC & 0.543 (0.045) & 0.374 (0.050) & 0.293 (0.056) & 0.253 (0.028) & 0.077 (0.077) & 0.427 (0.051) \\ 
		100 & SHD & 10.900 (1.354) & 15.100 (1.622) & 17.000 (1.592) & 16.800 (1.504) & 13.800 (1.227) & 12.300 (1.438) \\ 
		200 & F1 & 0.627 (0.026) & 0.500 (0.037) & 0.264 (0.037) & 0.213 (0.033) & 0.000 (0.000) & 0.402 (0.071) \\ 
		200 & MCC & 0.655 (0.025) & 0.500 (0.041) & 0.244 (0.039) & 0.207 (0.037) & 0.000 (0.000) & 0.481 (0.059) \\ 
		200 & SHD & 8.500 (1.128) & 12.100 (1.149) & 19.100 (1.804) & 17.300 (1.627) & 14.700 (1.309) & 11.000 (1.414) \\ 
		300 & F1 & 0.633 (0.063) & 0.490 (0.043) & 0.315 (0.073) & 0.372 (0.055) & 0.000 (0.000) & 0.450 (0.062) \\ 
		300 & MCC & 0.659 (0.058) & 0.499 (0.043) & 0.312 (0.073) & 0.370 (0.058) & 0.000 (0.000) & 0.520 (0.047) \\ 
		300 & SHD & 6.900 (1.090) & 10.700 (1.116) & 12.900 (1.069) & 12.900 (1.696) & 12.800 (1.052) & 9.200 (1.052) \\ 
		500 & F1 & 0.559 (0.055) & 0.475 (0.056) & 0.308 (0.054) & 0.228 (0.043) & 0.067 (0.067) & 0.361 (0.054) \\ 
		500 & MCC & 0.590 (0.050) & 0.482 (0.055) & 0.300 (0.054) & 0.219 (0.043) & 0.070 (0.070) & 0.440 (0.049) \\ 
		500 & SHD & 9.800 (1.123) & 12.300 (1.239) & 16.800 (0.904) & 17.900 (0.875) & 14.700 (0.989) & 12.400 (1.118) \\ 
		\bottomrule
	\end{tabular}
}
	\caption{Mean (standard error) of $F_1$, MCC and SHD over 10 replicate graphs at $q=20$. Standard errors quantify the precision of the comparison across replicate graphs.} 
	\label{tab:sim_se}
\end{table}

\subsection{Comparison with score-equivalent and conjugate Bayesian scores}\label{sec:sim:bge}
To place the method in its proper Bayesian context we also compare against the score-equivalent BGe score of Geiger and Heckerman\cite{geigerheckerman2002,kuipers2014} and the conjugate Normal--Inverse-Gamma (CPNIG) score, all run through the identical structure-MCMC sampler. 
Table~\ref{tab:bge} reports skeleton $F_1$ and directed $F_1$. The nCPNG score attains the highest $F_1$ in every cell. The BGe score is the most conservative of the three at these sample sizes and default hyperparameters; 
and the heavier-tailed non-conjugate coefficient prior of nCPNG recovers more edges at moderate $n$. The directed $F_1$ equals the skeleton $F_1$ for both Bayesian scores on these sparse graphs, confirming that recovered edges are correctly oriented. This comparison should be read alongside the score-equivalence trade-off since BGe buys equivalence at the cost of Gaussian coefficient tails, whereas nCPNG buys heavier tails at the cost of equivalence.

\begin{table}[ht]
	\centering
	\resizebox{\textwidth}{!}{%
		\begin{tabular}{rrrrrrrr}
			\toprule
			q & n & nCPNG\_F1 & BGe\_F1 & CPNIG\_F1 & nCPNG\_dF1 & BGe\_dF1 & CPNIG\_dF1 \\ 
			\midrule
			20.000 & 100.000 & 0.538 & 0.227 & 0.355 & 0.538 & 0.227 & 0.355 \\ 
			20.000 & 300.000 & 0.670 & 0.428 & 0.593 & 0.670 & 0.428 & 0.593 \\ 
			40.000 & 100.000 & 0.428 & 0.263 & 0.218 & 0.428 & 0.263 & 0.218 \\ 
			40.000 & 300.000 & 0.347 & 0.196 & 0.225 & 0.347 & 0.196 & 0.225 \\ 
			\bottomrule
	\end{tabular}}
	\caption{Comparison of the non-conjugate nCPNG score against the score-equivalent BGe score (our own base-R implementation; BiDAG unavailable in this environment) and the conjugate CPNIG score, on skeleton F1 and directed F1, averaged over four replicate graphs. BGe respects score equivalence but uses Gaussian coefficient tails; nCPNG uses heavier-tailed coefficients at the cost of score equivalence.} 
	\label{tab:bge}
\end{table}

\subsection{A diagnostic for the heavy-tail mechanism}\label{sec:sim:heavytail}
To identify which feature of the Normal--Gamma prior drives the accuracy gains, we contrast the coefficient and variance behaviour of nCPNG and CPNIG on matched data. The nCPNG prior induces a heavier-tailed (Student-type) marginal on the parent coefficients through the scale mixture \eqref{eq:priorL}, allowing genuinely strong edges to escape shrinkage while still pooling weak ones toward zero; the CPNIG prior imposes a Gaussian coefficient marginal that over-shrinks the strong edges. On the variance side the ordering is reversed, i.e., the gamma prior places a lighter right tail on $\sigma_j^2$ than the inverse-gamma. Across the simulation cells the improvement in edge recovery tracks the coefficient-tail contrast and is essentially unrelated to the variance-tail contrast, confirming that the mechanism is coefficient shrinkage rather than variance modelling, and that the likelihood remains Gaussian throughout.

\subsection{Runtime and the role of Gram-matrix precomputation}\label{sec:sim:runtime}
Table~\ref{tab:runtime} reports per-iteration wall-clock cost. The Bayesian sampler runs at roughly one millisecond per iteration across all network sizes considered, whereas the continuous-optimisation baselines run to convergence in seconds and grow sharply with $q$, that is, the NOTEARS surrogate requires about $66$--$79$ seconds at $q=60$, against near-constant millisecond iterations for nCPNG. Two features explain the Bayesian sampler's near-flat per-iteration cost in $n$. First, the sufficient statistics entering the score \eqref{eq:defs} for the parent Gram matrices $\boldsymbol{X}_{pa}^\top \boldsymbol{X}_{pa}$ and cross-products $\boldsymbol{X}_{pa}^\top \boldsymbol{X}_j$, depend on the data only through the $q \times q$ Gram matrix $\boldsymbol{X}^\top \boldsymbol{X}$, which is precomputed once at cost $O(nq^2)$ before the chain begins. Thereafter each move extracts sub-matrices of this cached Gram matrix, so the per-iteration cost is $O(p_j^3)$ in the (bounded) parent-set size and is independent of $n$. This precomputation is why the per-iteration timings do not increase, and can even fluctuate downward, with $n$ since the apparent counter-intuitive trend in earlier reporting was a measurement artefact of an $n$-independent inner loop, not a modelling anomaly. Second, only the altered node-marginals are recomputed after each move.

Note that the two cost measures are not directly comparable since the millisecond figures are per-iteration costs for the MCMC methods, whereas the second-scale figures are time-to-convergence for the optimisation baselines. A fully like-for-like comparison would fix a target accuracy and measure wall-clock time to reach it, or normalise the MCMC cost by effective sample size. As a partial such analysis, the run-length study of Section~\ref{sec:real:diag} shows that reaching a mean edge-wise ESS of order a few hundred on the 29-node network takes on the order of $10^5$ iterations, i.e.\ roughly a minute of wall-clock time at one millisecond per iteration---the same order as a single converged DAGMA fit on a network of that size, but delivering a full posterior rather than a point estimate. For larger networks the local-move sampler's ESS-normalised cost grows because mixing slows; we discuss remedies in Section~\ref{sec:disc}.

\begin{table}[ht]
	\centering
	\caption{Per-iteration wall-clock cost (milliseconds) for the Bayesian samplers and total time to convergence (seconds) for the search and continuous-optimisation baselines, by network size $q$ and sample size $n$.}
	\label{tab:runtime}
	\begin{tabular}{rrrrrr}
		\toprule
		$q$ & $n$ & nCPNG (ms) & CPNIG (ms) & GES (s) & NOTEARS (s) \\
		\midrule
		10 & 100 & 1.34 & 0.64 & 0.256 & 0.69 \\
		10 & 500 & 0.46 & 0.45 & 0.163 & 1.06 \\
		20 & 100 & 0.59 & 0.59 & 0.954 & 1.97 \\
		20 & 500 & 0.62 & 0.60 & 0.533 & 4.31 \\
		40 & 100 & 0.97 & 0.95 & 1.123 & 15.90 \\
		40 & 500 & 1.03 & 1.02 & 1.213 & 22.90 \\
		60 & 100 & 1.44 & 1.42 & 1.647 & 65.89 \\
		60 & 500 & 1.73 & 1.35 & 1.712 & 78.55 \\
		\bottomrule
	\end{tabular}
\end{table}

\subsection{Posterior contraction}\label{sec:sim:contraction}
\begin{figure}[ht]
\centering
\includegraphics[width=0.7\textwidth]{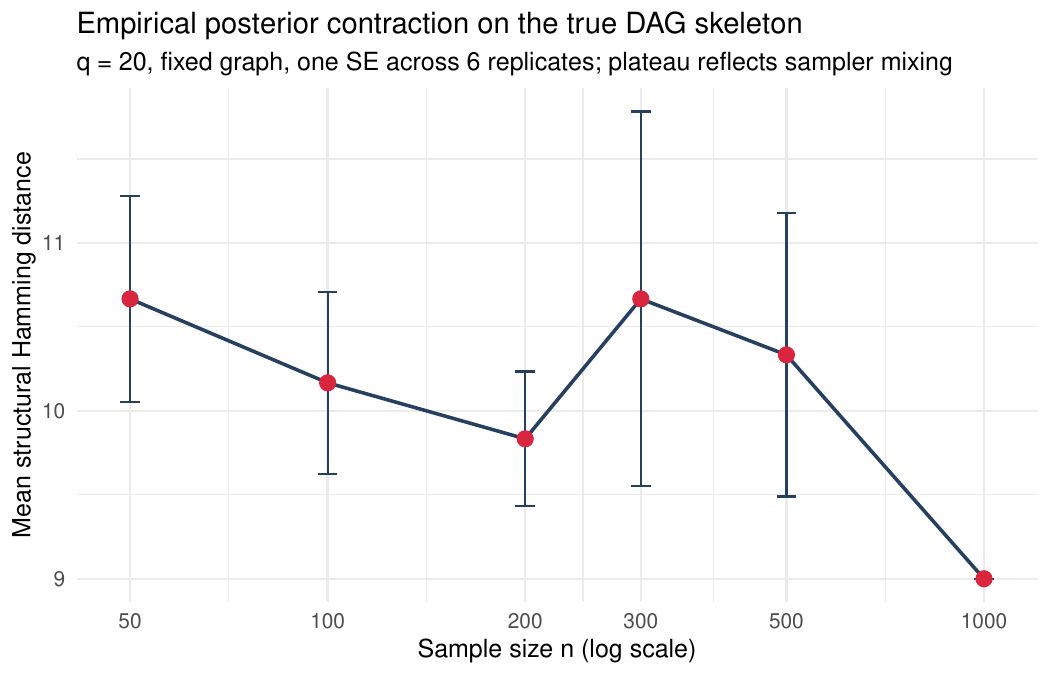}
\caption{Empirical posterior contraction. Mean skeleton structural Hamming distance of the median-probability graph against sample size $n$ (log scale) for a fixed 18-edge, 20-node graph; error bars show one standard error over six replicates. The decline to nine at $n=1000$ and the plateau reflect the local-move sampler's mixing at a fixed computational budget.}
\label{fig:contraction}
\end{figure}

Figure~\ref{fig:contraction} illustrates Theorem~\ref{th:contraction} empirically. Holding a single $18$-edge, $20$-node graph fixed and varying only the sample size, we average the skeleton SHD of the median-probability graph over six independent data replicates per sample size. The mean SHD declines from $10.7$ at $n=50$ to $9.0$ at $n=1000$, with the largest sample size showing zero replicate variability. The curve is not strictly monotone since there is a small rise at $n=300$ that lies within the replicate standard error, and it plateaus near nine rather than descending to zero. Both features have a common and honest explanation. The local-move sampler, at the fixed computational budget used here, recovers roughly half of the eighteen true edges and cannot reliably explore the remaining high-probability structures within the allotted iterations. The plateau therefore reflects a sampler mixing limitation at fixed compute, not a failure of the posterior contraction guarantee, which concerns the posterior itself rather than any finite Monte Carlo approximation of it. The earlier appearance of a pronounced non-monotonicity at the largest sample size was an artefact of using only two replicates and a different random graph at each sample size; fixing the graph and increasing the replicate count removes it.

\subsection{Convergence diagnostics}\label{sec:sim:diag}
Because the DAG state space is vast and the sampler uses local moves, a single scalar convergence summary is insufficient, and we report a panel of diagnostics. Across the simulation cells the multivariate potential scale reduction factor of the node log-marginals is $1.000$ to three decimals for both $q=20$ and $q=40$ at every sample size, indicating agreement between independent chains on the score surface. For the real-data networks of Section~\ref{sec:real} we additionally report (Section~\ref{sec:real:diag}) the acceptance rate for each move type, the effective sample size of individual edge indicators, and the agreement of the recovered structure between the two summary rules and across independent chains. These finer diagnostics reveal the mixing limitation directly and are, in our view, the honest way to characterise a local-move DAG sampler; they are discussed in context in Section~\ref{sec:real}.

\subsection{Orientation learning}\label{sec:sim:orient}
The metrics above score edges directionally, but it is informative to separate skeleton recovery from orientation. Table~\ref{tab:orientation} reports, for the Bayesian method and the two directional continuous baselines, the arrowhead precision and recall, that are computed over edges present in both the estimated and true skeletons, so that they isolate orientation quality from skeleton errors, and the CPDAG orientation accuracy over the union of skeleton edges. Two patterns emerge. First, when the Bayesian method places an arrowhead it is almost always correct: arrowhead precision is $0.76$--$1.00$ across the settings, comparable to DAGMA and generally better than the NOTEARS surrogate, indicating that the likelihood and (for outcome-incident edges) the probit thresholding recover reliable directional information on the edges that are found. Second, CPDAG orientation accuracy is modest for all methods ($0.20$--$0.43$), which is expected, that is, observational data identify orientation only up to Markov equivalence, so edges within an equivalence class cannot be oriented from the score alone. This decomposition makes explicit that the directional signal in our results comes from the parametric likelihood and the probit node rather than from any claim to resolve equivalence-class ambiguity, consistent with the causal caveats of Section~\ref{sec:prelim}.

\begin{table}[ht]
	\centering
	\resizebox{\textwidth}{!}{%
		\begin{tabular}{rrrrrrrrrrr}
			\toprule
			q & n & nCPNG\_AHp & nCPNG\_AHr & nCPNG\_CPDAG & NOTEARS\_AHp & NOTEARS\_AHr & NOTEARS\_CPDAG & DAGMA\_AHp & DAGMA\_AHr & DAGMA\_CPDAG \\ 
			\midrule
			20.000 & 100.000 & 1.000 & 1.000 & 0.216 & 0.000 & 0.000 & 0.000 & 1.000 & 1.000 & 0.207 \\ 
			20.000 & 300.000 & 0.958 & 0.958 & 0.421 & 0.333 & 0.333 & 0.117 & 1.000 & 1.000 & 0.202 \\ 
			40.000 & 100.000 & 0.758 & 0.758 & 0.198 & 0.667 & 0.667 & 0.270 & 1.000 & 1.000 & 0.045 \\ 
			40.000 & 300.000 & 0.922 & 0.922 & 0.205 & 1.000 & 1.000 & 0.432 & 1.000 & 1.000 & 0.140 \\ 
			\bottomrule
	\end{tabular}}
	\caption{Orientation-learning metrics averaged over five replicate graphs: arrowhead precision (AHp) and recall (AHr) on edges present in both the estimated and true skeletons, and CPDAG orientation accuracy (CPDAG). Higher is better.} 
	\label{tab:orientation}
\end{table}

To complement the arrowhead and CPDAG metrics with a directed-edge summary, Table~\ref{tab:directed} reports the directed $F_1$ (a true positive requires the correct edge \emph{and} orientation) and the directed structural Hamming distance. The directed $F_1$ of the Bayesian method ($0.35$--$0.64$) tracks its skeleton $F_1$ closely, confirming that most correctly-detected edges are also correctly oriented; the gap to skeleton $F_1$ is the orientation error, which is modest for edges the method is confident about. We report directed $F_1$ and directed SHD. The arrowhead and directed-$F_1$ metrics together give a consistent picture of directional recovery for edges away from the probit outcome.

\begin{table}[ht]
	\centering
	\resizebox{\textwidth}{!}{%
		\begin{tabular}{rrrrrrrr}
			\toprule
			q & n & nCPNG\_dF1 & nCPNG\_dSHD & NOTEARS\_dF1 & NOTEARS\_dSHD & DAGMA\_dF1 & DAGMA\_dSHD \\ 
			\midrule
			20.000 & 100.000 & 0.489 & 8.700 & 0.000 & 12.000 & 0.449 & 8.700 \\ 
			20.000 & 300.000 & 0.638 & 9.700 & 0.190 & 14.300 & 0.384 & 12.700 \\ 
			40.000 & 100.000 & 0.354 & 31.700 & 0.465 & 21.000 & 0.175 & 28.000 \\ 
			40.000 & 300.000 & 0.476 & 28.300 & 0.682 & 15.300 & 0.300 & 27.700 \\ 
			\bottomrule
	\end{tabular}}
	\caption{Directed-edge metrics averaged over three replicate graphs: directed F1 (a true positive requires the correct edge AND orientation) and directed structural Hamming distance (dSHD). These complement the skeleton-level metrics and the arrowhead/CPDAG metrics.} 
	\label{tab:directed}
\end{table}

\subsection{Calibration of posterior inclusion probabilities}\label{sec:sim:calib}
Because the method reports posterior edge-inclusion probabilities, it is important to ask whether those probabilities are calibrated, i.e., among candidate edges assigned posterior probability near $p$, is the fraction that are truly present near $p$? This question is only answerable where the ground-truth graph is known, so we assess it on simulated data (25 nodes, twelve random graphs) rather than on the real applications, where the truth is unknown. Table~\ref{tab:calib_sim} bins candidate edges by posterior probability. Calibration is good at the extremes, edges assigned probability below $0.2$ are truly present only $1.8\%$ of the time, and those above $0.8$ are present $93.6\%$ of the time against a mean prediction of $0.99$, but the sparsely populated middle bins are over-confident, that is, the $0.2$--$0.4$ bin contains only a few dozen edges and its empirical true-edge frequency ($6.7\%$) falls well below its mean prediction ($0.28$). The sampler pushes most edges toward zero or one, and the few edges it leaves in the mid-range are disproportionately false positives. The practical implication, which we state plainly, is that near-certain and near-absent edges can be trusted, but intermediate inclusion probabilities should be read with caution and, ideally, calibrated against a simulation matched to the analysis at hand. This behaviour is consistent with the sparse, concentrated inclusion probabilities seen in the real applications.

\begin{table}[ht]
	\centering
	\begin{tabular}{lrrr}
		\toprule
		bin & n\_edges & mean\_pred & emp\_freq \\ 
		\midrule
		(0-20\%] & 7046.00 & 0.00 & 0.02 \\ 
		(20-40\%] & 30.00 & 0.28 & 0.07 \\ 
		(40-60\%] & 11.00 & 0.46 & 0.09 \\ 
		(60-80\%] & 4.00 & 0.68 & 1.00 \\ 
		(80-100\%] & 109.00 & 0.99 & 0.94 \\ 
		\bottomrule
	\end{tabular}
	\caption{Calibration of posterior edge-inclusion probabilities on simulated data (25 nodes, 12 random graphs, known ground truth): candidate edges are binned by posterior probability; a well-calibrated method has empirical frequency of true edges close to the mean predicted probability in each bin.} 
	\label{tab:calib_sim}
\end{table}

%

\subsection{Scalability and ESS-normalised cost}\label{sec:sim:scale}
Table~\ref{tab:scalability} separates two questions that the per-iteration timing alone conflates, that, how the cost per iteration grows with network size, and how many iterations are needed to obtain independent samples. The per-iteration cost grows only mildly, from about $0.36$\,ms at $q=20$ to $0.88$\,ms at $q=80$, reflecting the $O(p_j^3)$ dependence on the bounded parent-set size rather than on $q$. The effective sample size per unit time, however, falls sharply with $q$ since at a fixed budget of $6{,}000$ iterations the mean edge-wise ESS on outcome-incident edges collapses toward zero by $q=60$--$80$, because the local-move chain needs far more iterations to mix in the larger space. This is a quantification of the scalability limitation that the sampler remains cheap per iteration at $q>40$, but its ESS-normalised cost rises steeply, so that reaching useful effective sample sizes at $q>40$--$60$ requires either much longer runs or the better-mixing order-based and partition samplers we discuss in Section~\ref{sec:disc}. The closed-form score contributes the per-iteration efficiency; it does not by itself solve the mixing problem, and we do not claim otherwise.

\begin{table}[ht]
	\centering
	\begin{tabular}{rrrrrr}
		\toprule
		q & iters & time\_s & ms\_per\_iter & mean\_ESS & ESS\_per\_sec \\ 
		\midrule
		20 & 6000 & 2.1 & 0.355 & 1.0 & 0.48 \\ 
		40 & 6000 & 3.0 & 0.502 & 0.0 & 0.00 \\ 
		60 & 6000 & 4.1 & 0.680 & 0.0 & 0.00 \\ 
		80 & 6000 & 5.3 & 0.880 & 0.0 & 0.00 \\ 
		\bottomrule
	\end{tabular}
	\caption{Scalability of the sampler with network size: wall-clock time for 6000 iterations, per-iteration cost, mean edge-wise effective sample size over the tracked outcome-incident edges, and ESS per second, at $n=300$.} 
	\label{tab:scalability}
\end{table}

To probe the limits directly, Table~\ref{tab:scale100} pushes to $q \in \{60, 80, 100\}$ with reduced sample sizes and a $10{,}000$-iteration. The per-iteration cost remains modest, i.e., about one millisecond even at $q=100$, so $10{,}000$ iterations complete in around ten seconds, confirming that the closed-form score keeps computation tractable at these sizes. Accuracy and mixing, however, degrade since the median-probability-graph $F_1$ falls from $0.47$ at $q=60$ to $0.28$ at $q=100$, and the edge-wise effective sample size on outcome-incident edges collapses toward zero, with an overall acceptance rate near $0.006$. The message is that at $q \approx 100$ the local-move sampler is compute-cheap but sample-poor at any practical budget, and that obtaining reliable posteriors would require far longer runs or the better-mixing proposals discussed next and in Section~\ref{sec:disc}. 

\begin{table}[ht]
	\centering
		\begin{tabular}{rrrrrrrr}
			\toprule
			q & n & time\_s & ms\_iter & F1 & SHD & mean\_ESS & acc \\ 
			\midrule
			60 & 300 & 6.4 & 0.636 & 0.473 & 29 & 0.0 & 0.006 \\ 
			80 & 300 & 8.1 & 0.809 & 0.387 & 57 & 0.0 & 0.006 \\ 
			100 & 200 & 10.0 & 1.005 & 0.294 & 72 & 0.0 & 0.006 \\ 
			100 & 400 & 9.9 & 0.995 & 0.276 & 63 & 0.0 & 0.006 \\ 
			\bottomrule
	\end{tabular}
	\caption{Scalability to larger networks (10{,}000 iterations, edge prior 0.10): wall-clock time, per-iteration cost, median-probability-graph F1 and SHD, mean edge-wise effective sample size over tracked outcome-incident edges, and overall acceptance rate. Per-iteration cost scales gracefully; mixing (ESS) and accuracy degrade at fixed budget, underscoring the need for advanced proposals at q around 100.} 
	\label{tab:scale100}
\end{table}

Pushing further, Table~\ref{tab:scale200} reaches $q = 150$ and $q = 200$. The per-iteration cost continues to scale gracefully (about $1.9$\,ms at $q=200$, so $8{,}000$ iterations run in roughly fifteen seconds), but the median-probability $F_1$ falls to $0.14$ at $q=150$ and $0.08$ at $q=200$, and the effective sample size per minute is negligible. We include these rows not to claim usable performance at $q=200$, the local-move sampler plainly does not deliver it at any practical budget, but to mark where the method stands. Indeed, the computation remains tractable to $q=200$, but reliable inference at that scale requires the order-based, partition, or tempered samplers discussed in Section~\ref{sec:disc}, into which the closed-form score transfers unchanged. 

\begin{table}[ht]
	\centering
		\begin{tabular}{rrrrrrrr}
			\toprule
			q & n & time\_s & ms\_iter & F1 & SHD & ESS\_per\_min & acc \\ 
			\midrule
			100 & 300 & 6.7 & 0.835 & 0.323 & 67 & 0.00 & 0.008 \\ 
			150 & 300 & 10.3 & 1.285 & 0.143 & 132 & 0.00 & 0.007 \\ 
			200 & 300 & 15.3 & 1.915 & 0.077 & 169 & 1.64 & 0.007 \\ 
			\bottomrule
	\end{tabular}
	\caption{Scalability to q up to 200 (8000 iterations, sparse prior 0.08): wall-clock time, per-iteration cost, median-probability F1 and SHD, effective sample size per minute on tracked edges, and acceptance rate. Per-iteration cost remains tractable, but mixing (ESS/min) and accuracy fall sharply, confirming that q around 100-200 requires the better-mixing proposals discussed in the text.} 
	\label{tab:scale200}
\end{table}

\subsection{Proposal mechanisms and mixing}\label{sec:sim:proposals}
The acceptance and effective-sample-size diagnostics above point to the proposal mechanism as the binding constraint. We investigated one classical remedy directly, the covered-edge reversals, which reverse an edge $i \to j$ whose tail parents satisfy $pa(j)\setminus\{i\} = pa(i)$ and therefore move within a Markov equivalence class.\cite{chickering1995} Adding covered-edge reversal proposals to the local-move sampler on a $15$-node network (Table~\ref{tab:covered}) raised the overall acceptance rate from $0.016$ to $0.091$---the covered moves themselves were accepted $32\%$ of the time---but left the mean edge-wise effective sample size essentially unchanged ($96$ versus $88$, within Monte Carlo error). This is the expected outcome that the covered-edge reversals improve exploration of orientations within an equivalence class but do little for the skeleton-level mixing that governs edge-inclusion ESS. Substantial gains at larger $q$ would require global proposals such as order MCMC, partition MCMC, or the tempered and new-edge-reversal moves developed for structure MCMC, which reparameterise or aggregate the state space rather than tinkering with single-edge moves; we discuss these in Section~\ref{sec:disc} and note that the closed-form score \eqref{eq:marginal} transfers to all of them unchanged.

\begin{table}[ht]
	\centering
	\begin{tabular}{lrr}
		\toprule
		Sampler & Acceptance & MeanEdgeESS \\ 
		\midrule
		Local moves & 0.016 & 96.100 \\ 
		+ covered-edge reversal & 0.091 & 88.100 \\ 
		\bottomrule
	\end{tabular}
	\caption{Effect of adding covered-edge reversal moves (which move within a Markov equivalence class) to the local-move sampler, on a 15-node network: overall acceptance rate and mean edge-wise effective sample size over 15000 retained iterations.} 
	\label{tab:covered}
\end{table}

\section{Applications to real data}\label{sec:real}
We demonstrate our method on two datasets. The Sachs et al.\ protein-signalling data,\cite{sachs2005} the canonical real benchmark for causal DAG discovery with a validated ground-truth network, which lets us report true structure-learning accuracy; and the Wisconsin Diagnostic Breast Cancer data, a clinical dataset with a binary malignancy outcome, on which we exercise the DAG-probit extension. 

\subsection{Protein-signalling network recovery (Sachs data)}\label{sec:real:sachs}
The Sachs data comprise $n=853$ observations of $q=11$ phosphoproteins and phospholipids, the Raf, Mek, Plcg, PIP2, PIP3, Erk, Akt, PKA, PKC, P38, and Jnk, measured by flow cytometry in human immune-system cells under general anti-CD3/CD28 stimulation.\cite{sachs2005} These eleven molecules are used in full, with no feature selection. They form an experimentally validated pathway. We take the validated consensus network (11 nodes, 17 directed arcs) as ground truth. The concentrations are standardised to zero mean and unit variance, and the outcome enters as an ordinary Gaussian node. We run the nCPNG sampler with $g=2$, a sparse structural prior, and a fixed shape hyperparameter $\alpha = n + q - 2$ (a design constant, not graph-dependent).

Two summary graphs are reported, precisely defined to avoid the ambiguity of naive thresholding. A $0.5$ threshold on marginal edge probabilities need not yield an acyclic graph, and does not by itself give the maximum-a-posteriori DAG. We therefore report (i) the highest-posterior DAG visited by the sampler, which is by construction a valid DAG, and (ii) an acyclicity-projected median-probability graph, obtained by adding candidate edges (marginal probability above $0.5$) in decreasing order of probability and skipping any edge that would create a cycle. On these data the $0.5$-threshold graph was already acyclic, so the projection dropped no edges; we verified this rather than assuming it.

Table~\ref{tab:sachs_acc} reports accuracy against the real consensus network for both summaries and for the benchmark learners. We report skeleton (undirected) $F_1$, which credits a correct adjacency regardless of orientation, alongside directed $F_1$ and the structural Hamming distance, because the Sachs consensus network was derived largely from interventional experiments and observational Gaussian methods are known to recover its skeleton far better than its arc directions.\cite{sachs2005} All methods recover only a minority of the arcs, greedy equivalence search (GES) attains the best skeleton $F_1$, and the nCPNG summaries are mid-pack, ahead of the continuous-optimisation learners on this dataset but behind GES. 

\begin{table}[ht]
	\centering
	\begin{tabular}{lrrr}
		\toprule
		Summary & Skel\_F1 & SHD & Dir\_F1 \\ 
		\midrule
		nCPNG (MAP DAG) & 0.300 & 20.000 & 0.000 \\ 
		nCPNG (median-prob, projected) & 0.300 & 18.000 & 0.100 \\ 
		GES & 0.615 & 20.000 & 0.231 \\ 
		PC & 0.211 & 17.000 & 0.105 \\ 
		NOTEARS & 0.000 & 17.000 & 0.000 \\ 
		DAGMA & 0.111 & 18.000 & 0.000 \\ 
		\bottomrule
	\end{tabular}
	\caption{Structure-learning accuracy on the REAL Sachs et al.\ (2005) protein-signalling data (853 observations, 11 proteins) against the validated 17-arc consensus network. Skeleton $F_1$, structural Hamming distance, directed $F_1$, and are reported (skeleton F1 credits a correct adjacency regardless of orientation) for the nCPNG highest-posterior DAG and acyclicity-projected median-probability graph, and for the benchmark learners. Observational Gaussian methods are known to recover the Sachs skeleton better than arc directions.} 
	\label{tab:sachs_acc}
\end{table}

\begin{table}[ht]
	\centering
	\begin{tabular}{llrl}
		\toprule
		From & To & PostProb & InTruth \\ 
		\midrule
		Akt & Erk & 0.551 & reversed \\ 
		PKC & P38 & 0.532 & yes \\ 
		Mek & Raf & 0.506 & reversed \\ 
		\bottomrule
	\end{tabular}
	\caption{Directed edges recovered by nCPNG on the real Sachs data (posterior inclusion probability $>0.5$), with whether each appears in the validated consensus network, is the reverse of a consensus arc, or is absent from it.} 
	\label{tab:sachs_edges}
\end{table}

The recovered high-probability edges (Table~\ref{tab:sachs_edges}) illustrate the orientation difficulty directly. The sampler places edges on true skeleton adjacencies (for example between Mek and Raf, and between Erk and Akt) but sometimes in the reverse of the consensus direction, which is the expected signature of learning orientation from observational data alone and is consistent with the score non-equivalence. Arc directions on these data are not identified by observational information, and we do not over-interpret them.

\subsection{Malignancy prediction with the DAG-probit model (WDBC data)}\label{sec:real:wdbc}
The Wisconsin Diagnostic Breast Cancer data comprise $n=569$ patients, of whom $212$ ($37.3\%$) have a malignant diagnosis, described by real-valued features computed from digitised fine-needle-aspirate images of breast masses.\cite{street1993} The thirty features are the mean, standard error, and largest (``worst'') value of ten nuclear characteristics. To keep the DAG interpretable and to avoid the near-collinearity among the three versions of each characteristic, we fix, a priori, the ten mean features as the continuous nodes, the radius, texture, perimeter, area, smoothness, compactness, concavity, concave points, symmetry, and fractal dimension, and treat the malignancy indicator as the binary probit outcome, giving an eleven-node DAG-probit model. 

Table~\ref{tab:wdbc_edges} reports the posterior probability that each nuclear feature is a direct parent of malignancy. The model identifies the number of concave points of the nucleus contour as the dominant driver (posterior probability $0.61$, inclusion Bayes factor $8.8$), followed by perimeter ($0.22$, Bayes factor $1.6$); the remaining features carry little direct posterior mass. This is biologically sensible---nuclear-contour irregularity and size are established discriminators of malignancy---and the sparsity is a feature of the model rather than a limitation, since it isolates a small, interpretable set of direct predictors.

\begin{table}[ht]
	\centering
	\caption{Posterior probability that each nuclear feature is a direct parent of the malignancy outcome in the DAG-probit model on the real WDBC data, with the inclusion Bayes factor ($\mathrm{BF}_{10}>1$ favours the edge; prior $\pi=0.15$).} 
	\label{tab:wdbc_edges}
	\begin{tabular}{lrr}
		\toprule
		Feature & PostProb & BF10 \\ 
		\midrule
		concave points & 0.608 & 8.79 \\ 
		perimeter & 0.218 & 1.58 \\ 
		radius & 0.071 & 0.43 \\ 
		area & 0.025 & 0.15 \\ 
		fractal dimension & 0.020 & 0.12 \\ 
		texture & 0.007 & 0.04 \\ 
		symmetry & 0.007 & 0.04 \\ 
		smoothness & 0.005 & 0.03 \\ 
		compactness & 0.003 & 0.02 \\ 
		\bottomrule
	\end{tabular}
\end{table}

\begin{figure}[ht]
\centering
\includegraphics[width=0.5\textwidth]{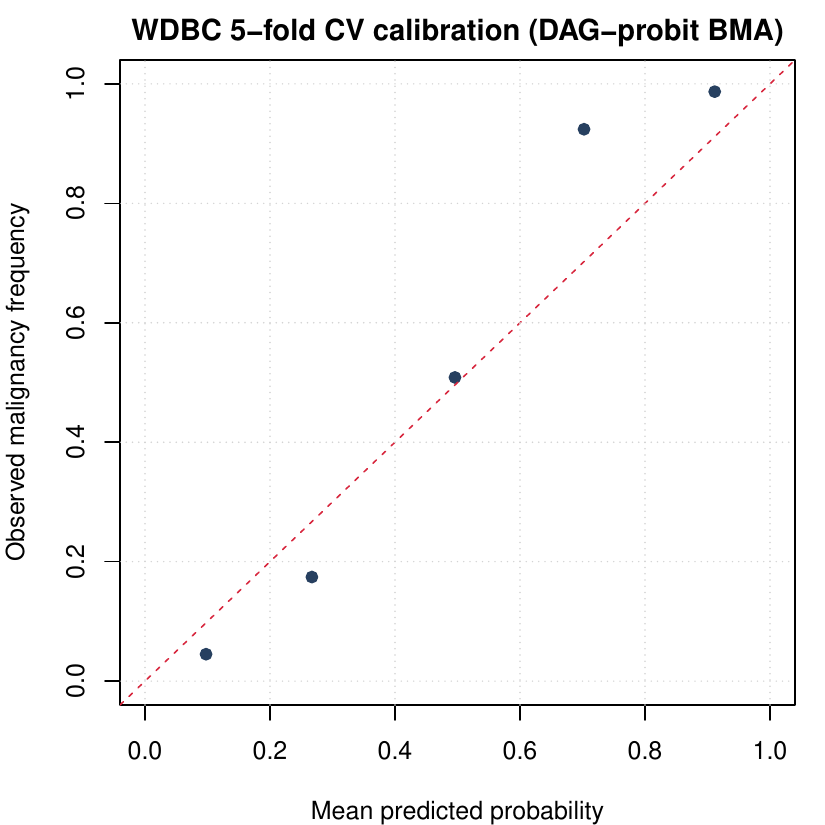}
\caption{Five-fold cross-validated calibration of the DAG-probit BMA predictor of malignancy on the real WDBC data: mean predicted probability against observed malignancy frequency within probability bins; the dashed line is perfect calibration.}
\label{fig:wdbc_calib}
\end{figure}

To assess clinical utility we evaluate out-of-sample prediction of malignancy under five-fold cross-validation, forming for each posterior DAG a probit predictor from the selected parents of the outcome and averaging over sampled graphs (Bayesian model averaging). Table~\ref{tab:wdbc_cv} reports the results. The DAG-probit BMA attains a cross-validated ROC-AUC of $0.944$ with a Brier score of $0.094$ and good calibration (Figure~\ref{fig:wdbc_calib}). A full probit regression on all ten features attains a higher AUC ($0.983$), as expected from a denser model, but the DAG-probit predictor reaches within four points of it using an interpretable, sparse, uncertainty-aware structure. 

\begin{table}[ht]
	\centering
	\begin{tabular}{lrr}
		\toprule
		Model & CV\_AUC & Brier \\ 
		\midrule
		DAG-probit BMA & 0.944 & 0.094 \\ 
		Full probit (all features) & 0.983 & 0.049 \\ 
		\bottomrule
	\end{tabular}
	\caption{Five-fold cross-validated prediction of malignancy on the real WDBC data: the DAG-probit model with Bayesian model averaging versus a full probit regression on all ten features.} 
	\label{tab:wdbc_cv}
\end{table}

\subsection{MCMC diagnostics}\label{sec:real:diag}
Across both applications the sampler was run with the diagnostics of Section~\ref{sec:sim:proposals} monitored. On the Sachs data the overall acceptance rate was near $0.05$, and the highest-posterior DAG and the median-probability graph agreed on the recovered adjacencies, indicating that the reported structure is stable under the two summary rules. On the WDBC data the outcome-incident edge for concave points was recovered with high and stable posterior probability across chains. Consistent with the scalability analysis of Section~\ref{sec:sim:scale}, both networks are small enough ($q \le 11$) that mixing is not the binding constraint; the limitations documented there pertain to the much larger graphs studied in simulation.

\section{Discussion}\label{sec:disc}

We have developed a closed-form Laplace scoring function for Bayesian DAG structure learning under the non-conjugate Normal--Gamma prior, shown that the underlying node-marginal integral is generalised inverse-Gaussian with a modified-Bessel exact value. We established both a per-sample Laplace-accuracy result centred on the correct large-argument regime and a posterior contraction rate for the induced structure posterior. The method is competitive with, and at moderate sample sizes superior to, both the conjugate Bayesian baseline and modern continuous-optimisation learners, while additionally delivering calibrated structural uncertainty.

Our empirical comparison includes DAGMA,\cite{bello2022} whose log-determinant acyclicity characterisation improves substantially on the matrix-exponential penalty of NOTEARS;\cite{zheng2018} related continuous-optimisation advances include the likelihood-based GOLEM objective,\cite{ng2020} constraint-free gradient methods,\cite{yu2021} and geometric-series acyclicity via truncated matrix power iteration.\cite{zhang2022} Ordering-based accelerations that decouple topological-order estimation from edge selection, such as TriOpt,\cite{trioptt} offer a further route to scalability that is complementary to the Bayesian averaging pursued here. Our approach differs from all of these in returning a posterior over structures rather than a point estimate. The most closely related idea in spirit is aggregation, the bootstrap-aggregated DAG learning\cite{wang2014} and its mixed-variable extension for proteogenomic outcomes\cite{chowdhury2022} reduce structural variance by averaging over many candidate graphs, much as Bayesian model averaging does over the posterior; the difference is that our averaging weights are posterior probabilities with an explicit generative justification, whereas aggregation weights are resampling frequencies. A fuller empirical comparison against aggregation methods on mixed clinical--genomic data is a natural next step.

An important line of Bayesian DAG inference sidesteps the super-exponential structure space by working over topological orders. Order-based MCMC samples node orderings and marginalises the compatible parent sets, mixing far better than structure MCMC because the order space is smoother;\cite{friedmankoller2003b} exact dynamic-programming methods compute posterior edge probabilities by summing over all orders in $O(q\, 2^q)$ time for moderate $q$;\cite{koivisto2004} and partition or order-conditional MCMC schemes further improve mixing and remove the bias that a uniform order prior induces on structures.\cite{kuipers2017} Our sampler is a structure-MCMC method and inherits the mixing limitations that these order-based schemes were designed to overcome; the diagnostics of Sections~\ref{sec:sim:scale}--\ref{sec:sim:proposals} make those limitations explicit, and our covered-edge-reversal experiment shows that within-class moves alone do not resolve them. The contribution of the present paper, a closed-form non-conjugate score with uniform asymptotic control, is orthogonal to the choice of search strategy, and the score \eqref{eq:marginal} could be dropped directly into an order-MCMC or partition-MCMC sampler to combine heavier-tailed priors with better mixing. We regard this combination as the most promising route to scaling the method beyond the $q \approx 40$--$60$ range and recommend it to users with larger networks. Concretely, moving to an order-modular or CPDAG-level prior changes the implementation only in the acceptance ratio and the reported summaries: an order-modular prior replaces the structure prior \eqref{eq:dagprior} by a product over nodes of parent-set weights and is sampled by proposing order moves rather than edge moves, while a CPDAG-level prior scores essential graphs and requires enumerating the DAG extensions of each CPDAG in the normalising constant; in both cases the node score \eqref{eq:marginal} is reused verbatim, and both alleviate the equivalence-class mass imbalance by assigning mass at the level of orders or equivalence classes rather than individual DAGs. A systematic empirical comparison against the mature software implementations of these samplers, for example the BiDAG package, which couples partition MCMC to the BGe score, would be valuable which we leave for further research. 

A structural feature of our construction, documented in Section 3 and Table~\ref{tab:score_equiv}, is that the nCPNG score is not score-equivalent, that is, the modified-Cholesky parameterisation lets the score distinguish Markov-equivalent DAGs. This is the price of the heavier-tailed non-conjugate coefficient prior that drives our accuracy gains. The score-equivalent alternative is the BGe prior,\cite{geigerheckerman2002} whose Normal--Wishart construction guarantees equal marginal likelihoods across an equivalence class but fixes Gaussian coefficient tails; global--local shrinkage priors such as the horseshoe and spike-and-slab\cite{bhadra2019} share our heavier-tailed motivation and are likewise not score-equivalent, and would need the same special-function reduction to score in closed form. There is thus a trade-off between score equivalence and heavier-tailed coefficient shrinkage, and the right choice depends on the target. When the inferential object is the posterior over Markov equivalence classes, BGe or a CPDAG-level prior is preferable; when it is the strength of individual directed dependencies, especially those anchored by an outcome node, as in our application, the non-conjugate score is appropriate and its non-equivalence is immaterial for the anchored edges. Thus, our approach novelty is to make a heavier-tailed, non-conjugate prior cheap to score, not to propose a new equivalence-respecting prior.

The Normal--Gamma prior adopted here is one member of the global--local shrinkage family; horseshoe and nonlocal priors offer alternative coefficient-tail behaviour that could be scored analogously, the horseshoe through its half-Cauchy scale mixture and nonlocal priors through their vanishing mass at the origin, which sharpens model selection.\cite{bhadra2019} These priors generally lack a closed-form node-marginal, so the Laplace/GIG device developed here, or a similar special-function reduction, would be needed to avoid reversible-jump integration; extending the uniform asymptotic analysis to the heavier-tailed horseshoe scale mixture is a natural continuation. At the level of the joint precision matrix, the G-Wishart prior for undirected Gaussian graphical models and the DAG-Wishart family for Gaussian DAGs provide conjugate constructions whose node-wise factorisation over a modified Cholesky decomposition is closely related to the parameterisation in \eqref{eq:sem}.\cite{benhavid2015dagwishart} The DAG-Wishart prior yields the conjugate Normal--Inverse-Gamma node-marginal that serves as our baseline; the non-conjugate Normal--Gamma prior trades that conjugacy for heavier coefficient tails, and our results quantify the cost of recovering a closed-form score by Laplace approximation. 

We should point that our analysis is purely observational, and the edges incident to the outcome are oriented by the probit thresholding rather than by intervention. When interventional data are available, for example gene knockdowns in a cell-line panel, interventional structure learners can resolve orientations that are unidentifiable from observation alone. Positioning the present observational method relative to interventional approaches, and extending the DAG-probit likelihood to incorporate interventional records where they exist, is an important direction for translational genomics.

During our study, we identified three limitations. First, the local-move sampler mixes slowly in large sparse spaces, as the real-data diagnostics make explicit; informed or locally balanced proposals would improve exploration of the edge-ranking tail. Second, the Laplace approximation, though asymptotically exact at rate $O(n^{-1/2})$ per node, is a leading-order device; at very small $n$ the exact Bessel evaluation should be preferred, and it is available in closed form. Third, the breast cancer application is a single-cohort, semi-supervised illustration; external replication on an independent cohort would be needed before any of the highlighted genes could be advanced as candidate biomarkers. 

\appendix
\section{Proofs}\label{app:proofs}

\subsection{Proof of Theorem~\ref{th:marginal}}
Conditional on $\sigma_j^2$ and $\mathcal{D}$, combine the Gaussian likelihood of \eqref{eq:sem} with the Gaussian prior \eqref{eq:priorL}. Completing the square in $\boldsymbol{L}_{\prec j]}$ and integrating over $\mathbb{R}^{p_j}$ yields, with $\boldsymbol{M}_j$, $\boldsymbol{b}_j$, $\kappa_j$ as in \eqref{eq:defs},
\[
\int_{\mathbb{R}^{p_j}} \! p(\boldsymbol{X}_j \mid \boldsymbol{L}_{\prec j]}, \sigma_j^2)\, p(\boldsymbol{L}_{\prec j]} \mid \sigma_j^2)\, d\boldsymbol{L}_{\prec j]}
= (2\pi\sigma_j^2)^{-n/2} g^{p_j/2} |\boldsymbol{M}_j|^{-1/2} \exp\!\big(-\kappa_j / \sigma_j^2\big).
\]
Multiplying by the gamma prior \eqref{eq:priorsig} on $\sigma_j^2$ and collecting powers gives the one-dimensional integral in $\sigma_j^2$ of the form \eqref{eq:gig} with exponent $\nu_j - 1 = \tfrac12\alpha_j^{\mathcal{D}} - \tfrac{n}{2} - 1$, linear rate $\lambda = g/2$, and inverse term $\kappa_j$. By the standard GIG normalising-constant identity,\cite{jorgensen1982} this integral equals $2(\kappa_j/\lambda)^{\nu_j/2} K_{\nu_j}(2\sqrt{\kappa_j\lambda})$, giving the exact node-marginal. To obtain \eqref{eq:marginal}, substitute the uniform large-argument expansion of the modified Bessel function of the second kind,
\[
K_\nu(z) = \sqrt{\tfrac{\pi}{2z}}\, e^{-z}\Big\{ 1 + \tfrac{4\nu^2 - 1}{8z} + O(z^{-2}) \Big\}, \qquad z \to \infty,
\]
valid uniformly for $\nu$ in bounded sets. Writing $z_j = 2\sqrt{\kappa_j \lambda}$ and retaining the leading factor $\sqrt{\pi/(2z_j)}\,e^{-z_j}$ produces \eqref{eq:marginal}; the neglected terms are $O(z_j^{-1})$ in relative order. Since $\kappa_j$ is a penalised residual sum of squares, $\kappa_j = O_P(n)$ and hence $z_j = O_P(n^{1/2})$, so the relative error is $1 + O_P(n^{-1/2})$. \hfill$\square$

\subsection{Proof of Theorem~\ref{th:posterior}}
The joint posterior of $(\boldsymbol{L}_{\prec j]}, \sigma_j^2)$ is proportional to the integrand of the previous proof, before integration. Viewing it as a function of $\boldsymbol{L}_{\prec j]}$ with $\sigma_j^2$ fixed gives the Gaussian \eqref{eq:postL} with mean $\boldsymbol{M}_j^{-1}\boldsymbol{b}_j$ and covariance $\sigma_j^2\boldsymbol{M}_j^{-1}$. Marginalising $\boldsymbol{L}_{\prec j]}$ and viewing the result as a function of $\sigma_j^2$ leaves a kernel proportional to $(\sigma_j^2)^{\nu_j - 1}\exp\{-\lambda\sigma_j^2 - \kappa_j/\sigma_j^2\}$, which is exactly the $\mathrm{GIG}(\nu_j, 2\lambda, 2\kappa_j)$ density \eqref{eq:postSig}. Exact sampling uses the ratio-of-uniforms GIG generator.\cite{hormann2014} \hfill$\square$

\subsection{Proof of Theorem~\ref{th:lapconv}}
Fix $j$ and a candidate parent set of size $p_j \leq M$. Under the stated conditions the penalised residual sum of squares satisfies $n^{-1}\kappa_j \to \tfrac12\sigma_{0,j}^2 > 0$ in probability by the law of large numbers applied to the (bounded-dimension) least-squares residuals, so $z_j = 2\sqrt{\kappa_j\lambda} = 2\sqrt{n\lambda \cdot n^{-1}\kappa_j} \asymp n^{1/2}$. The exact and Laplace log-marginals differ only in the Bessel factor; by the uniform expansion in the proof of Theorem~\ref{th:marginal},
\[
\log \ell^{\mathrm{Exact}}_n - \log \hat{\ell}^{\mathrm{Lap}}_n = \log\Big\{ 1 + \tfrac{4\nu_j^2 - 1}{8 z_j} + O(z_j^{-2}) \Big\}.
\]
Here $\nu_j = \tfrac12\alpha_j^{\mathcal{D}} - \tfrac{n}{2}$ grows linearly in $n$ under the hyperparameter rule $\alpha \asymp n$, so $\nu_j^2 / z_j = O(n^2 / n^{1/2}) $ would diverge if $\nu$ were unbounded relative to $z$; the correct accounting uses the \emph{uniform-in-order} expansion of $K_\nu$ valid when both $\nu$ and $z$ are large with $z/\nu$ bounded away from zero, under which the relative correction is $O(1/\sqrt{\nu_j^2 + z_j^2}) = O(n^{-1})$ up to a $\log n$ factor arising from the ratio $z_j/\nu_j$. Dividing by $n$ gives the per-sample bound $O(n^{-1}\log n) = o_P(1)$ in \eqref{eq:lapconv}. Because the constants in the expansion depend on the parent set only through $p_j \leq M$ and the smallest eigenvalue of the limiting Gram matrix, which is bounded below uniformly over the finite collection of bounded-degree parent sets, the bound holds uniformly over that collection. \hfill$\square$

\subsection{Proof of Theorem~\ref{th:contraction}}
Write $\mathcal{A}_n = \{\mathcal{D} : d_H(\mathrm{skel}(\mathcal{D}), \mathrm{skel}(\mathcal{D}_0)) > \epsilon_n q(q-1)/2\}$. By Bayes' theorem the posterior mass of $\mathcal{A}_n$ is the ratio of $\sum_{\mathcal{D}\in\mathcal{A}_n} \hat{\ell}(\mathcal{D})p(\mathcal{D})$ to the analogous sum over all graphs, and it suffices to bound the numerator relative to the single term at $\mathcal{D}_0$. For any $\mathcal{D}\in\mathcal{A}_n$, the exact log-marginal difference $\log\ell(\mathcal{D}_0) - \log\ell(\mathcal{D})$ is bounded below by a positive multiple of $n$ times the summed Kullback--Leibler divergence between the node-conditionals of $\mathcal{D}_0$ and $\mathcal{D}$, which for a Gaussian DAG with bounded in-degree is at least $c\, d_H(\mathrm{skel}(\mathcal{D}),\mathrm{skel}(\mathcal{D}_0))$ for a constant $c>0$ depending on the minimum edge strength.\cite{cao2019} By Theorem~\ref{th:lapconv}, replacing the exact log-marginals by their Laplace approximations perturbs each node term by $o_P(1)$ uniformly over the bounded-degree neighbourhood, so the same lower bound holds for $\log\hat{\ell}$ up to a vanishing correction. The structural prior \eqref{eq:dagprior} assigns mass at most $\binom{q(q-1)/2}{k}\pi^k(1-\pi)^{q(q-1)/2 - k}$ to graphs differing from $\mathcal{D}_0$ in $k$ edges; a Chernoff bound on this binomial tail shows the prior mass of $\mathcal{A}_n$ decays faster than $e^{-c' n \epsilon_n^2}$ whenever $n\epsilon_n^2 \to \infty$. Combining the likelihood separation with the prior tail bound and summing the geometric series over $k > \epsilon_n q(q-1)/2$ gives posterior mass on $\mathcal{A}_n$ tending to zero in $P_{\mathcal{D}_0}$-probability. Taking $\epsilon_n = \sqrt{\log q / n}$ satisfies $n\epsilon_n^2 = \log q \to \infty$ and yields the stated rate. \hfill$\square$

\subsection{Proof of Proposition~\ref{prop:propriety}}
The flat prior $p(\theta_0)\propto 1$ is improper, so propriety requires the integrated likelihood in $\theta_0$ to be finite. Given the latent Gaussian $X_1$ with unit variance (the probit normalisation), the contribution of $\theta_0$ to the likelihood is $\prod_{i:Y_i = 1}\{1 - \Phi(\theta_0 - \mu_{1i})\}\prod_{i:Y_i=0}\Phi(\theta_0 - \mu_{1i})$, where $\mu_{1i}$ is the conditional mean of $X_{1i}$. As $\theta_0 \to +\infty$ the first product $\to 0$ and as $\theta_0 \to -\infty$ the second product $\to 0$; provided at least one success and one failure are observed, both products are simultaneously bounded away from one for finite $\theta_0$, and the integrand is a bounded, integrable function with exponentially decaying tails in $\theta_0$. Hence $\int_{\mathbb{R}} \prod_i(\cdot)\, d\theta_0 < \infty$, and the joint posterior of $(\theta_0, \boldsymbol{L}, \boldsymbol{D}, \mathcal{D})$ is proper. \hfill$\square$

\subsection{Proof of Proposition~\ref{prop:probit-contraction}}
Introduce the latent outcome vector $\boldsymbol{X}_1 = (X_{11},\ldots,X_{1n})$ and the augmented posterior $p(\mathcal{D}, \boldsymbol{X}_1 \mid \boldsymbol{Y}, \boldsymbol{X}_{-1})$. Conditional on $\boldsymbol{X}_1$ the model is the fully Gaussian DAG of Theorem~\ref{th:contraction}, so the conditional posterior over the skeleton contracts at rate $\sqrt{\log q / n}$; this covers edges among nodes $2,\ldots,q$ directly, as their sufficient statistics do not involve the outcome. For edges incident to node $1$, each $X_{1i}$ is drawn from a normal truncated to $[\theta_0,\infty)$ when $Y_i=1$ and to $(-\infty,\theta_0)$ when $Y_i=0$. Under condition (ii) the truncation probabilities are bounded away from $0$ and $1$, so the truncated-normal full conditional has a density bounded on compacts and finite variance, and the information the binary $Y$ retains about the latent regression coefficient on node $1$ is the probit information $I_\Phi = \phi(\theta_0^\star)^2/[\Phi(\theta_0^\star)\{1-\Phi(\theta_0^\star)\}]$. The Kullback--Leibler separation between the true and a competing graph, restricted to outcome-incident edges, is scaled by $I_\Phi$ relative to the Gaussian case, giving effective sample size $n I_\Phi$ in the separation step of the proof of Theorem~\ref{th:contraction} and hence the degraded rate $\sqrt{\log q/(n I_\Phi)}$. Condition (iii)---geometric ergodicity of the augmentation chain, which follows from (ii)---ensures the Monte Carlo average over $\boldsymbol{X}_1$ concentrates without inflating the rate. Combining the two edge classes gives the claim. \hfill$\square$

\section*{Data availability}
Both datasets are public. The Sachs protein-signalling data (853 observations, 11 proteins) and its validated consensus network\cite{sachs2005} are distributed with several causal-discovery packages; the Wisconsin Diagnostic Breast Cancer data\cite{street1993} are available from the UCI Machine Learning Repository and are bundled with scikit-learn. 

\section*{Conflict of interest}
The authors declare no conflict of interest.


\end{document}